\documentclass[12pt,letterpaper]{article}   
\usepackage{osajnl} 

\usepackage{amssymb}
\usepackage{amsmath}
\usepackage{latexsym}
\usepackage{rotating}

\newcommand{\ds}{\displaystyle}
\newcommand{\du}{\mathrm{d}}

\newcommand{\mtotK}{\mathrm{E}_{K}}
\newcommand{\meps}{\mathrm{E}_{\epsilon}}

\begin{document}

\title{Fiber optic interferometry: \\
Statistics of visibility and closure phase}

\author{E. Tatulli, A. Chelli}

\affiliation{Laboratoire d'Astrophysique, Observatoire de Grenoble, 38041
  Grenoble cedex France}

\email{Eric.Tatulli@obs.ujf-grenoble.fr} 

\begin{abstract}Interferometric observations with three telescopes or
  more provide two observables:  closure phase
  information together with visibilities measurements. 
When using single-mode interferometers, both
  observables have to be redefined in the light of the coupling phenomenon between the incoming wavefront
  and the fiber. We introduce in this paper the estimator of both 
  so-called modal visibility and modal closure phase. Then,
  we compute the
  statistics of the two observables in presence of partial correction
  by Adaptive Optics, paying attention on the correlation
  between the measurements. We find that the correlation coefficients are mostly zero and
in any case
  never overtakes $1/2$ for the visibilities, and $1/3$ for the closure
  phases. From this theoretical analysis, data
  reduction process using classical least square minimization is investigated.
  In the framework of the
  AMBER instrument, the three beams recombiner of the VLTI, we simulate
  the observation of a single Gaussian source and we study the
  performances of the interferometer in terms of diameter
  measurements. We show that the observation is optimized, i.e. that
  the Signal to Noise Ratio (SNR) of the diameter is maximal, 
  when the full width half maximum (FWHM) of the source is roughly
  $1/2$ of the mean resolution of the interferometer. 
We finally point out that in the case of an observation with 3 telescopes, neglecting the correlation between the
measurements leads to overestimate the SNR by a factor of $\sqrt{2}$. 
We infer that in any cases, this value is an upper limit. 
\end{abstract}

\ocis{030.6600, 030.7060, 070.6020, 070.6110, 120.3180}

\maketitle


\section{Introduction}
Thanks to the simultaneous recombination of the light
  arising from three telescopes, interferometers such as IONIC-3T on
  IOTA\cite{berger_1} or AMBER on the VLTI\cite{petrov_1}
  are providing closure phase measurements together with the modulus of the visibility. 
Retrieval of phase information allows to scan the geometry of the
  source, hence opening the era of image reconstruction with infrared interferometric observations.  
  However, with the current number of telescopes 
  available, direct image restoration requires many successive nights of
  observations\cite{thiebaut_1}. Thus, in most of the cases where
  the ($u,v$) coverage spans  a relatively small number of spatial
  frequencies, the measurements have still to be analyzed in
the light of model-fitting techniques. 

Furthermore, together with
  partial correction by Adaptive Optics (AO), many of the
  up-to-date interferometers are making use of waveguides that
  spatially filters
  the atmospheric corrugated wavefront, changing the turbulent phase
  fluctuations into random intensity variations\cite{foresto_2} . 
  The estimators that
  describes the visibility and the closure phase measurements
  obtained
  with such interferometers have to account for the coupling
  between the partially corrected wavefront and the fiber. 
  From these appropriate estimators one can derive
  the statistical properties of the observables and
  properly investigate the performances of single-mode interferometers.

In Section \ref{sec_theory} we recall the
  spatial filtering properties of the waveguides in terms of
  interferometric signal and we
  define the estimators of both the modal visibility and the closure phase. We investigate in Section
  \ref{sec_noise} the covariance matrices of the observables with respect to atmospheric,
  photon and detector noises, paying particular attention to the
  correlation coefficients. Then, defining in
  Section \ref{sec_modelfit} a general least square model fitting
  of the measurements, we analyze in Section \ref{sec_applications}
  the ability of fiber optic interferometers to measure
  stellar diameters.


\section{Principles of fiber optic interferometry} \label{sec_theory}
\begin{figure}[!*p]
\begin{center}
\includegraphics[width=16cm]{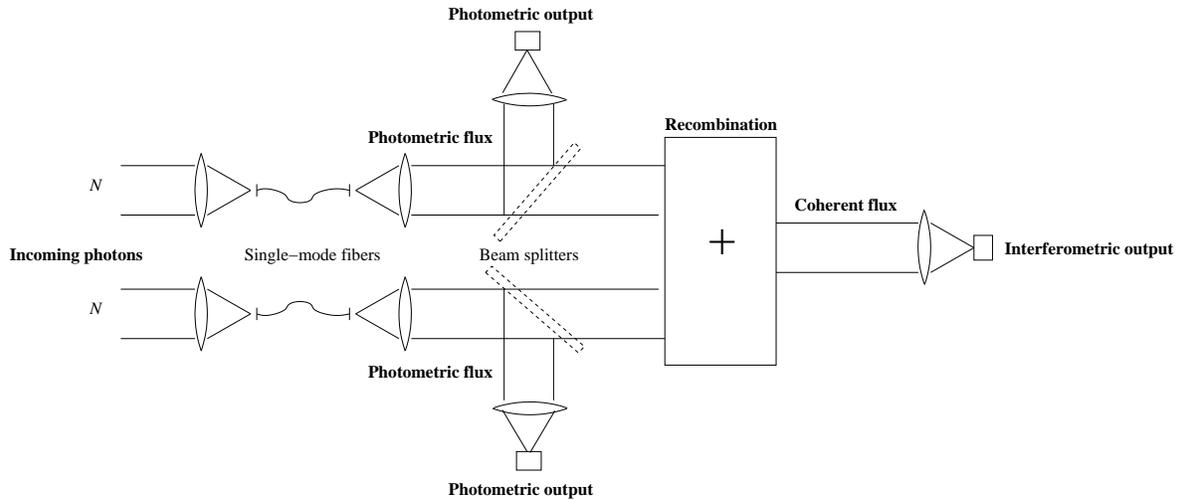} 
\caption{\label{fig_recombineur} Sketch of a fiber optic
  interferometer. Thanks to single mode fibers, beams coming for each
  telescope are spatially filtered. Then for each beam, 
  a beam splitter is selecting
  one fraction of the light in order to estimate the photometric flux
  through dedicated photometric output. The remaining part of the
  light is recombined with the other beams (recombination is seen here
  as a black box, and the recombination mode -- spatial or temporal --
  is not a relevant parameter in this study). At the output of the
  recombination, the coherent flux is estimated thanks to the so-called
  interferometric output.}
\end{center}
\end{figure}

A full analysis of the signal arising from fiber optic 
interferometers has
been  theoretically described by M\`ege\cite{mege_1} and summarized
by Tatulli et al\cite{tatulli_1} . 
We only recall here the important points for
this paper, focusing on the coupling phenomenon between the incoming
wavefront and the fiber, and on the observables that can be obtained from
such interferometers. Figure \ref{fig_recombineur}
  sketches the principle of a fiber optic interferometer, and reports
  the main technical terms that will be used all along this paper.    

\subsection{Spatial filtering}
Introducing waveguides to carry/recombine the light in an
interferometer allows to perform a spatial filtering of the incoming
wavefront. It means that the phase corrugation of the wavefront are
changed into intensity fluctuations. In other words, the number of
photometric and coherent (interferometric) photoevents at
the output of the fibers depends on atmospheric fluctuations and
results in the coupling between the turbulent wavefront and the
fibers\cite{foresto_2}$^,$\cite{ruilier_1} . 
Hence, spatial filtering can be seen
as coupling coefficients, i.e. the fraction of (respectively
photometric and coherent) light that is captured by the
fibers. Such coupling coefficients are mathematically described by the
following equations\cite{dyer_1}$^,$\cite{mege_4} :
\begin{equation}
\rho_i(V_{\star}) = \rho_0 {(V_{\star}*T^i)_{f=0}}
\label{eq_rhoi}  
\end{equation}
\begin{equation}
\rho_{ij}(V_{\star}) = \rho_0 (V_{\star}*T^{ij})_{f=f_{ij}} \label{eq_rhoij}
\end{equation}
where $V_{\star}$ is the visibility of the source and $T^{i}$ and $T^{ij}$ 
are resulting in respectively the auto-correlation and
cross-correlation of the aberration-corrupted pupil weighted
by the fiber single mode\cite{mege_4}$^,$\cite{roddier_1}~:
\begin{eqnarray}
T^{i}(f) &=& \frac{\int P_{i}(r)P_{i}^{\ast}(r+\lambda{f})
  \psi_i(r)\psi_i^{\ast}(r+\lambda{f})\du{r}}{\int |P_{i}(r)|^2
  \du{r}} \\
T^{ij}(f) &=& \frac{\int P_{i}(r)P_{j}^{\ast}(r+\lambda{f})
  \psi_i(r)\psi_j^{\ast}(r+\lambda{f})\du{r}}{\int P_{i}(r)P_{j}^{\ast}(r)
  \du{r}} 
\end{eqnarray}
where $P_{i}(r)$ is the pupil function of the $i^{th}$ fiber optic 
telescope and
 $\psi_i(r)$ is the aberration-corrupted wavefront incoming on the $i^{th}$ pupil.  
$T^{i}$ and $T^{ij}$ are respectively called the photometric and
interferometric peaks. 
The inverse Fourier transform of $T^{i}$ is called the photometric
 lobe (or antenna lobe as commonly named in radio) and the inverse
 Fourier transform of $T^{ij}$  is called the interferometric lobe.
$\rho_0$
is the optimum coupling efficiency 
fixed by the fiber core design\cite{shaklan_1} . 

In the case where the source is unresolved by a single telescope
(i.e. is much tighter than the photometric lobe),
Eq. \ref{eq_rhoi} can be simplified:
\begin{equation}
\rho_i(V_{\star}) = \rho_0 \int T^i(f) \du{f} = \rho_0 \mathcal{S}^i \label{eq_rhoi_simplified}
\end{equation}
where $\mathcal{S}$ is the instantaneous Strehl ratio\cite{foresto_1} .
Moreover, if the 
visibility is constant over the range of the high
frequency peak $T^{ij}$, the interferometric coupling coefficient has
also a simplified expression:
\begin{equation}
|\rho_{ij}(V_{\star})|^2 = \rho_0^2 \mathcal{S}^i\mathcal{S}^j|
 V_{\star}(f_{ij})|^2 \label{eq_rhoij_simplified}
\end{equation}
Under these conditions, the effect of spatial filtering by the fibers in the
interferometric equation and, as 
a result, in the observables, is entirely characterized by the
instantaneous Strehl ratio statistics. 

\subsection{Estimation of the modal visibility}
We refer to Tatulli et al\cite{tatulli_1} 
for a more complete description of the
modal visibility. Note however that the expression of the coherent
flux at the spatial frequency $f_{ij}$ is given by:
\begin{equation}
I(f_{ij}) = \sqrt{K_iK_j} \rho_{ij}(V_{\star}) \label{eq_fluxcoherent}
\end{equation}
where $K_i$ and $K_j$ are the number of photoevents of telescopes $i$
and $j$ before entering the fiber.
An estimator of the modal visibility in the Fourier 
space is given by the ratio of the coherent flux by the photometric
ones, assuming the latter are estimated independently through dedicated 
photometric outputs (see Fig. \ref{fig_recombineur}):
\begin{equation}
\widetilde{V_{ij}^2}
=\frac{<|I(f_{ij})|^2>}{<k_ik_j>}\left(\frac{\tau}{1-\tau}\right )^2 \label{eq_v
is_estimate}
\end{equation}
$\tau$
is the fraction of light selected for photometry at the
output of the beam-splitter, and $k_i$, $k_j$ are the photometric
fluxes  (after the fibers).
\subsection{Modal bispectrum and closure phase}
By definition, the closure phase is the phase of the so called image bispectrum $\widetilde{B}_{klm}$. The latter
consists in the ensemble average of the triple product
\makebox{$<\widehat{I}(f_{kl})\widehat{I}(f_{lm})\widehat{I}^{\ast}(f_{km})>$}.
It can be expressed from Eq. \ref{eq_fluxcoherent} as:
\begin{eqnarray}
&&\widetilde{B}_{klm}=
<K_kK_lK_m\rho_{kl}(V_{\star})\rho_{lm}(V_{\star})\rho^{\ast}_{km}(V_{\star})>\\
&&= K_kK_lK_m\rho_0^3\int\!\!\!\!\!\int\!\!\!\!\!\int V_{\star}(f)V_{\star}(f^{'})
V^{\ast}_{\star}(f^{''}). \nonumber \\
&&<T^{kl}(f_{kl}-f)T^{lm}(f_{kl}-f^{'})T^{km^{\ast}}(f_{kl}-f^{''})> \du{f}\du{f^{'}}\du{f^{''}}\label{eq_bispectrum_est}
\end{eqnarray}
Using Roddier's formalism\cite{roddier_2} that demonstrated 
the bispectrum analysis to be a
generalization to the optical of the well known phase closure method
currently used in radio interferometry, it is straightforward to
notice that the quantity
$<T^{kl}(f)T^{lm}(f^{'})T^{km^{\ast}}(f^{''})>$ is non zero if
$f^{''} = f + f^{'}$ and that in this case:
\begin{equation}
<T^{kl}(f)T^{lm}(f^{'})T^{km^{\ast}}(f^{''})> =
\frac{N(f,f^{'})}{N^3(0)} 
\end{equation}
\begin{equation}
 K_kK_lK_m V_{\star}(f)V_{\star}(f^{'})
V^{\ast}_{\star}(f^{''}) = B_{\star}(f,f^{'})
\end{equation}
where $N(f,f^{'})$ is proportional to the overlap area of
three pupil images shifted apart by the spacings $f$, $f^{'}$
and $f^{''} = f + f^{'}$, and $B_{\star}(f,f^{'})$ is the bispectrum of 
the source. 
Hence the modal bispectrum can be rewritten:
\begin{equation}
\widetilde{B}_{klm} = \rho_0^3\int\!\!\!\!\!\int
B_{\star}(f,f^{'})  \frac{N(f_{kl}-f,f_{lm}-f^{'})}{N^3(0)}
\du{f}\du{f^{'}} \label{eq_modalbis}
\end{equation}
Thus, the modal bispectrum arising 
from fiber optic interferometers is the source bispectrum integrated over
the overlap area $N(f_{kl}-f,f_{lm}-f^{'})$.
 Hence, as the modal visibility does not equal in general the object visibility, 
the modal bispectrum does not coincide with the source bispectrum. 
Nevertheless, if the
source spectrum is constant over the overlap area $N(f_{kl}-f,f_{lm}-f^{'})$, Eq. \ref{eq_modalbis} takes a simplified form:
\begin{equation}
\widetilde{B}_{klm} = \rho_0^3B_{\star}(f_{kl},f_{lm})  \int\!\!\!\!\!\int
\frac{N(f,f^{'})}{N^3(0)} \du{f}\du{f^{'}} \label{eq_bispec}
\end{equation}
In this case, the modal bispectrum is proportional to that of the source.


\section{The covariance matrices}\label{sec_noise}
\begin{figure}[!*p]
\begin{center}
\begin{tabular}{lccr}
\includegraphics[width=5cm,height=4cm]{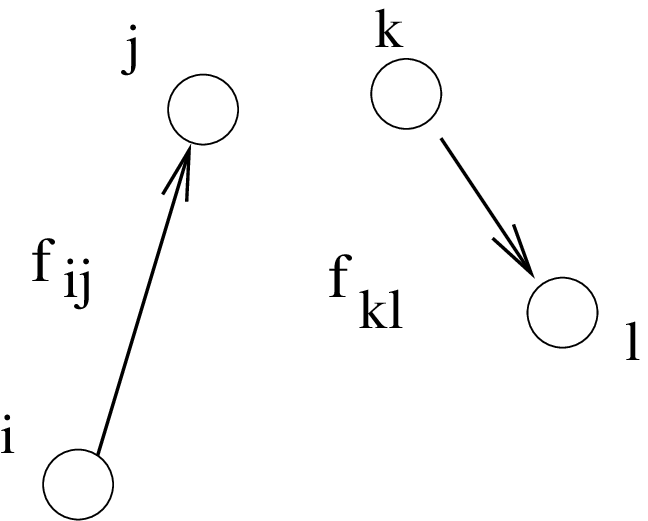} &&& \includegraphics[width=4cm,height=4cm]{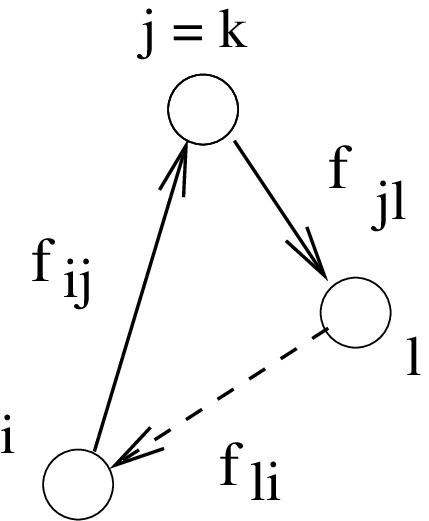}
\end{tabular}
\caption{\label{fig_vis_config}This figure shows the 
configurations that have to be 
    considered when computing the covariance of the visibilities. 
Respectively no telescope in common (left) or one telescope in
    common, hence forming a closure (right).}
\end{center}
\end{figure}

\begin{figure}[!*p]
\begin{center}
\begin{tabular}{cc}
\includegraphics[width=6cm]{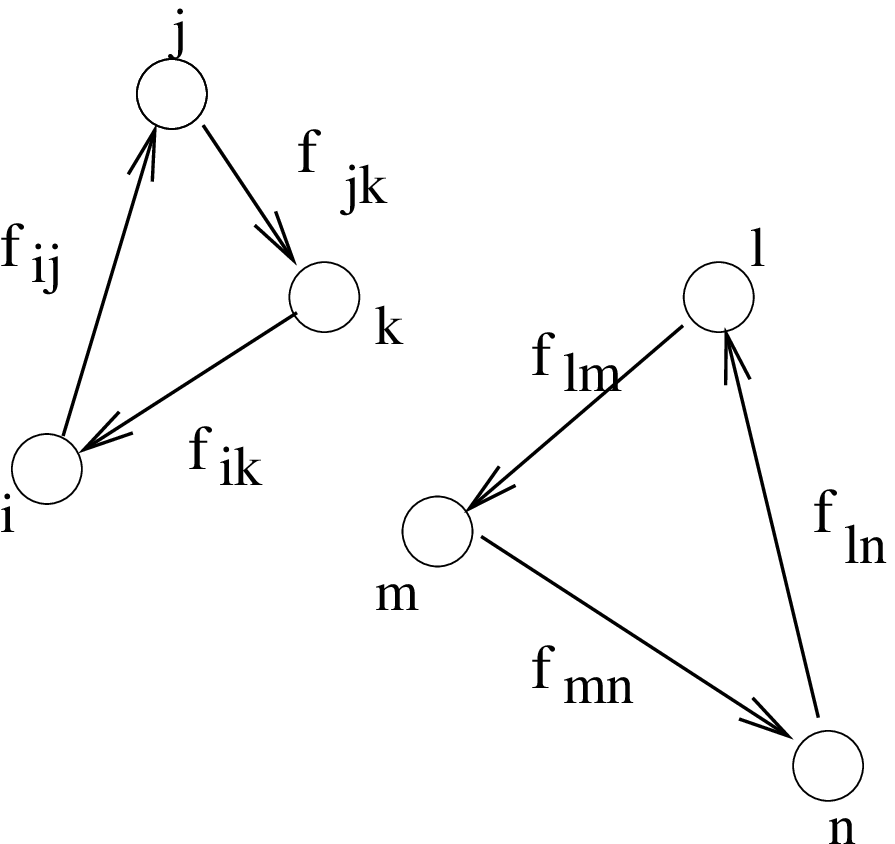} & \includegraphics[width=5.8cm]{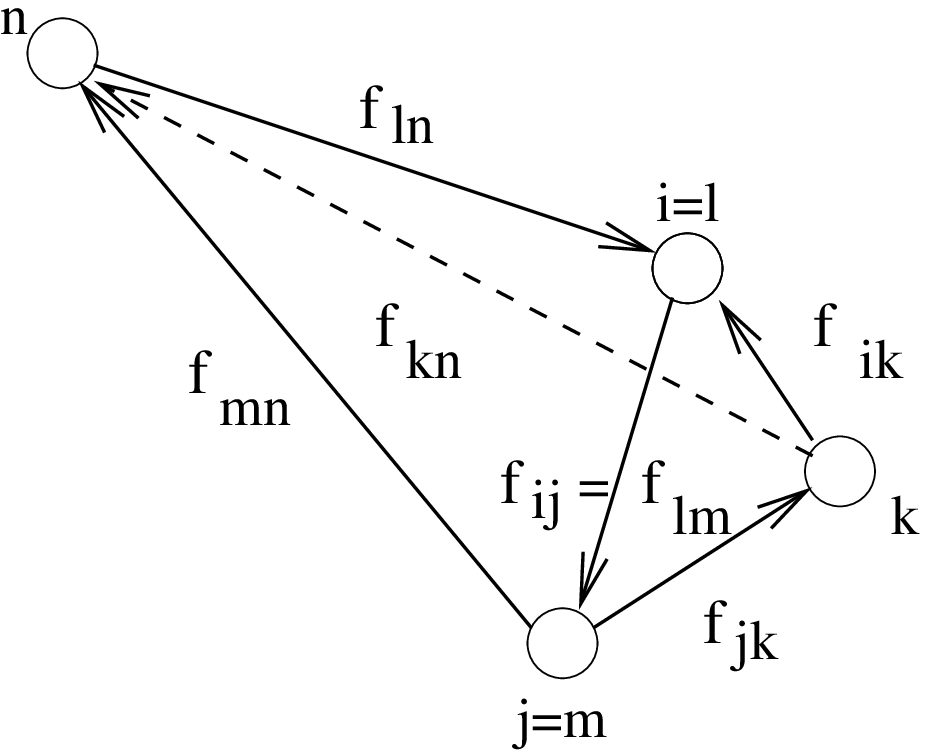}
\end{tabular}
\caption{\label{fig_cov_config}This figure shows
    configurations that have to be  
    considered when computing the covariance of the closure
    phases. Respectively no baseline (left) or one baseline (right) in
    common between both triplets of telescopes.}
\end{center}
\end{figure}

We propose to characterize the statistics of the square
visibility and that of the closure phase (i.e. the bispectrum phase) 
by computing their respective covariance matrix. Our objective is twofold:  
derive the error associated to each observable and
investigate the degree of dependency of each observable through 
their correlation coefficients. In order to do so, we use
the spatially continuous model of
photodetection introduced by Goodman\cite{goodman_1}$^,$\cite{chelli_1}
where the
signal is corrupted by three different types of noise: 
(i) the signal photon noise; (ii) the
additive Gaussian noise of global variance $\sigma^2$
which arises from the detector and from thermal emission; (iii) the atmospheric
noise resulting from the coupling efficiency variations due to the
turbulence. To simplify the calculations, we
assume that the source is unresolved by a 
single aperture, such that the low frequency coupling coefficients
verify Eq. \ref{eq_rhoi_simplified}, 
and that the source visibility is constant over the range of the interferometric peaks, 
such that the high frequency coupling coefficients verify Eq. \ref{eq_rhoij_simplified}. 
These assumptions drive to neglect the modal speckle
noise regime --  it has been shown that the modal
speckle noise is rejected towards negative magnitudes and only
  affects very bright sources\cite{tatulli_1} -- and to only focus on ``photon noise" and
``detector noise" regimes . 
The full calculations of the covariance coefficients of the
visibilities and of the closure phases 
are done in Appendix A. 
They lead to relatively complicated formulae which depend on
the Strehl statistics.
 Using a simple analytical approach, we derive in Appendix Bthe mean 
and the variance of the Strehl as a function of the turbulence strength and the level of AO 
correction. The relative error of the Strehl is bounded between two
limit values\cite{roddier_1}$^,$\cite{goodman_1} : 

\begin{equation}
\mathrm{(Perfect~correction)}~~  0 \le \frac{\sigma_{\mathcal{S}}}{\overline{ \mathcal{S}}} \le 1 ~~\mathrm{(No~correction)}
\end{equation} 
Table \ref{table_error}. gives the expressions of the 
limiting values of the variance of the visibility
and the closure phase for a point source in both ``photon noise"
and ``detector noise" regimes.
The error of the visibility will be deeply used in the next section 
to derive the performances of the Very Large Telescope Interferometer
(VLTI) with regards to single sources diameter measurements. 
Let us concentrate in this part
on the correlation coefficients of the visibilities and the closure phases, respectively. Their limiting values are
summarized in Table \ref{table_noises}. 
Clearly, for visibilities the correlation coefficients are null in the photon noise regime when no telescope is in 
common (see Fig. \ref{fig_vis_config}.) and are always smaller than $1/2$
otherwise, and for the closure phases they are null when no baseline
is in common (see Fig. \ref{fig_cov_config}.) 
and are always smaller than $1/3$ otherwise. 
Furthermore, the relative number of null elements 
in the covariance matrices rapidly increases 
with the number of telescopes. It implies that
if there are $(N-1)(N-2)/2$ {\it linearly independent} closure phase
relations\cite{monnier_1} , the whole set of $N(N-1)(N-2)/6$ closure phase relations 
can be considered in first approximation as {\it statistically independent}. 

It would be of great interest to pursue this study by comparing  
  the statistics of the closure phase to that of the phase
  measured by phase referencing technique. The formalism presented in
  Appendix A  can indeed 
  be transposed to the phase referencing case, but
  it requires a dedicated analysis which is beyond the scope of
  the present paper.   


\section{Model fitting} \label{sec_modelfit}

The present generation of interferometers only provides small number of telescopes (basically $2$ or $3$). In such a case, it is frequent 
that the lack of spatial frequencies in the $(u,v)$ coverage prevents from image reconstruction of the 
studied object. Hence, the measurements have to be analyzed in the light of a  model of the object. We propose here a simple $\chi^2$ 
model-fitting of the observables.

Let us define $\widehat{O}_{\theta}$ as the normalized spectrum of the  object model characterized by a set of parameters  $\theta$. 
From this spectrum 
we derive the model of the observables that have to be fitted according to the measurements:\\

\begin{tabular}{ll}
Square visibility & $\ds |\widehat{O}_{\theta}|^2(f_{ij})$ \\ 
Closure Phase & $\ds \phi_{G} =  \arctan\left[\frac{\mathrm{Im}(G_{\theta})}{\mathrm{Re}(G_{\theta})}\right]$
\end{tabular}\\ 

\noindent 
where $ G_{\theta}^{klm} = \widehat{O}_{\theta}(f_{kl})\widehat{O}_{\theta}(f_{lm})\widehat{O}_{\theta}^{\ast}(f_{km})$ is the bispectrum model.
Then, the estimated parameters $\widetilde{\theta}$ constraining at
best the observations are obtained by minimizing the distance between
the model and the measurements. Assuming Gaussian statistics for the observables, the
distance is given by the well known $\chi^2$:
\begin{equation}
\chi^2_{tot} = \chi^2_{V^2} + \chi^2_{\phi_{B}} \label{eq_chi2_tot}
\end{equation}
with
\begin{equation}
\chi^2_{V^2} =
\left[\widetilde{V^2}-|\widehat{O}_{\theta}|^2\right]C_{V^2}^{-1}\left[\widetilde{V^2}-|\widehat{O}_{\theta}|^2\right]^{T}
\label{eq_chi2_vis} 
\end{equation}
\begin{equation}
\chi^2_{\phi_{B}} = \left[\widetilde{\phi}_{B} - \phi_{G}\right]C_{\phi_{B}}^{-1}\left[\widetilde{\phi}_{B} - \phi_{G}\right]^{T} \label{eq_chi2_closephase}
\end{equation}
where $X^T$ denotes the transpose of the vector $X$. Equation
\ref{eq_chi2_tot} supposes that the two
observables of different nature (i.e. the visibility and the closure
phase) are uncorrelated. Such assumption seems quite reasonable since
we have shown that two measurements of same nature are 
already poorly correlated.

It can be argued that, in order to avoid phase discontinuities issues, it
might be better to work on the phasor (i.e. the average bispectrum) than on the
phase of the average bispectrum. 
In that case the $\chi^2$ constraint takes the
form:
\begin{equation}
\chi^2_{B} = \left[\mathrm{Im}\left(\widetilde{B}G_{\theta}^{\ast}\right)\right]C_{\mathrm{Im}\left(\widetilde{B}G_{\theta}^{\ast}\right)}^{-1}\left[\mathrm{Im}\left(\widetilde{B}G_{\theta}^{\ast}\right)\right]^{T}
\end{equation}
Such constraint does not appear to be
appropriate for two reasons:  (i) constraining the average
bispectrum or the closure phase of the average bispectrum is
strictly equivalent when the latter shows good SNR;
(ii) the covariance depends on the model which
makes the $\chi^2$ minimization subject to biases due to improper noise estimates. 
For very noisy data, the closure phase shows a lot of
discontinuities and its probability law, wrapped around $[-\pi,\pi]$, 
tends towards a uniform
law. Such a case, for which the $\chi^2$ fitting method is no more
optimal, neither for the closure phase nor for the bispectrum,
corresponds in practice to the sensitivity limit of the instrument.

Putting  Equation \ref{eq_chi2_tot} into the generic form:
\begin{equation}
\chi^2(\theta) =\mathcal{M}_{\theta} \mathcal{C}_{\mathcal{M}}^{-1} \mathcal{M}_{\theta}^{T}
\end{equation}
the error of the estimated parameters $\widetilde{\theta}$
writes\cite{heide_knoechel_1} :
\begin{equation}
\sigma^2(\widetilde{\theta}) =
\mathrm{Diag}\left\{\left[\mathcal{A}^T\mathcal{C}_{\mathcal{M}}^{-1}\mathcal{A}\right]^{-1}\right\} \label{eq_error_param}
\end{equation}
where 
\begin{equation}
\mathcal{A} = \frac{\partial \mathcal{M}_{\theta}}{\partial \theta} \bigg|_{\theta=\widetilde{\theta}}
\end{equation}

\section{Applications} \label{sec_applications}
\subsection{Observing a single Gaussian source}
We propose in this section to simulate the
 observation of a single Gaussian source of FWHM $\sigma_{O}$, with
 AMBER, the three beam recombiner of the VLTI\cite{petrov_1} . Since the object is centro-symmetrical, 
closure phase is not relevant and hence all the information is contained in the visibility alone. For sake
of simplicity, we first neglect the contribution of the correlation coefficients. Their effect will be 
studied in Section \ref{sec_correlations}. We adopt the standard 
instrumental configuration of AMBER\cite{malbet_1} in which the signal is sampled over $N_{pix}=16$ pixels. 
We choose  a spectral
resolution of $35$ in the $K$ band ($2.2\mu m$), an integration time of $30\mathrm{ms}$ per interferogram, a
transmission coefficient $\tau = 0.5$, a readout noise of $15\mathrm{e}^{-}/\mathrm{pix}$ and an optimized coupling
coefficient of $\rho_0=0.8$\cite{shaklan_1} .
We observe an object with $2$ Unit
Telescopes (UT2 and UT4, $D=8\mathrm{m}$) during $4\mathrm{H}$ (half of the time on
the object, half of the time on the calibrator)
with $5\mathrm{min}$ per frequency point which, together with the
 integration time per interferogram, leads to $5000$ samples per frequency
point. Note that all along the observation, the length of the
 projected baseline 
is quite constant with $B=45\mathrm{m} \pm 2\mathrm{m}$. 
Finally, we assume  a
turbulence strength of $D/r_0=5$ 
and a typical AO correction of $\mathcal{S}=0.5$.

\begin{figure}[!*p]
 \begin{center}
 \begin{tabular}{c}
 \includegraphics[width=12cm]{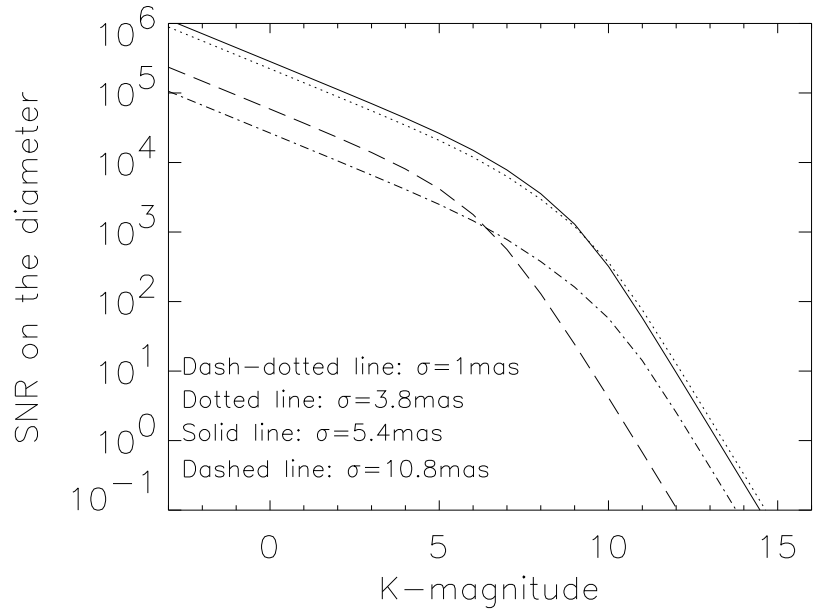}
 \end{tabular}
 \caption{\label{fig_snr_diam_K}SNR of the diameter 
 $\sigma_o$ as a function of the magnitude. Different sizes are considered:
 $\sigma_o=1\mathrm{mas}$ (dash-dotted line) ,
 $\sigma_o=3.8\mathrm{mas}$ (dotted line),
 $\sigma_o=5.4\mathrm{mas}$ (solid line), $\sigma_o=10.8\mathrm{mas}$
 (dashed line). The resolution is $\lambda/B =
 10\mathrm{mas}$. $\mathcal{S}=0.5$, $D/r_0=5$.}
 \end{center}
 \end{figure}

Fig. \ref{fig_snr_diam_K}. shows the SNR of the object size
 as the function of the magnitude. We consider different
sizes: $\sigma_o=1\mathrm{mas}$,
 $\sigma_o=3.8\mathrm{mas}$, $\sigma_o=5.4\mathrm{mas}$, and
 $\sigma_o=10.8\mathrm{mas}$, which are to be compared to $\lambda/B \sim 10\mathrm{mas}$, the resolution of the interferometer. 
As expected, a general
profile can be seen with two well known regimes: the ``photon
noise" regime for bright sources and  the ``detector noise" regime for
 faint sources. Defining the limiting magnitude as the magnitude 
 for which the SNR is equal to $1$, we find $\mathcal{K} \sim 11 - 13$ according to the size
 of the source. 

Clearly, the SNR first increases and then decreases with the size of the source, reaching a maximum
 around $\sigma_o=4\mathrm{mas}$. This phenomenon can be understood as follows: for marginally resolved sources, 
the parameters of the fit are barely constrained and the SNR is small. It increases with the size up to a point 
where the available projected baseline range does not match anymore the frequency content of the object. From there, the SNR begins 
to drop. This trend stands for all observing conditions, but as we show in the next section the exact position 
of the maximum depends on the noise regime.  

\subsection{Optimizing the baseline}
\begin{figure}[!*p]
 \begin{center}
 \begin{tabular}{c}
  \includegraphics[width=8cm]{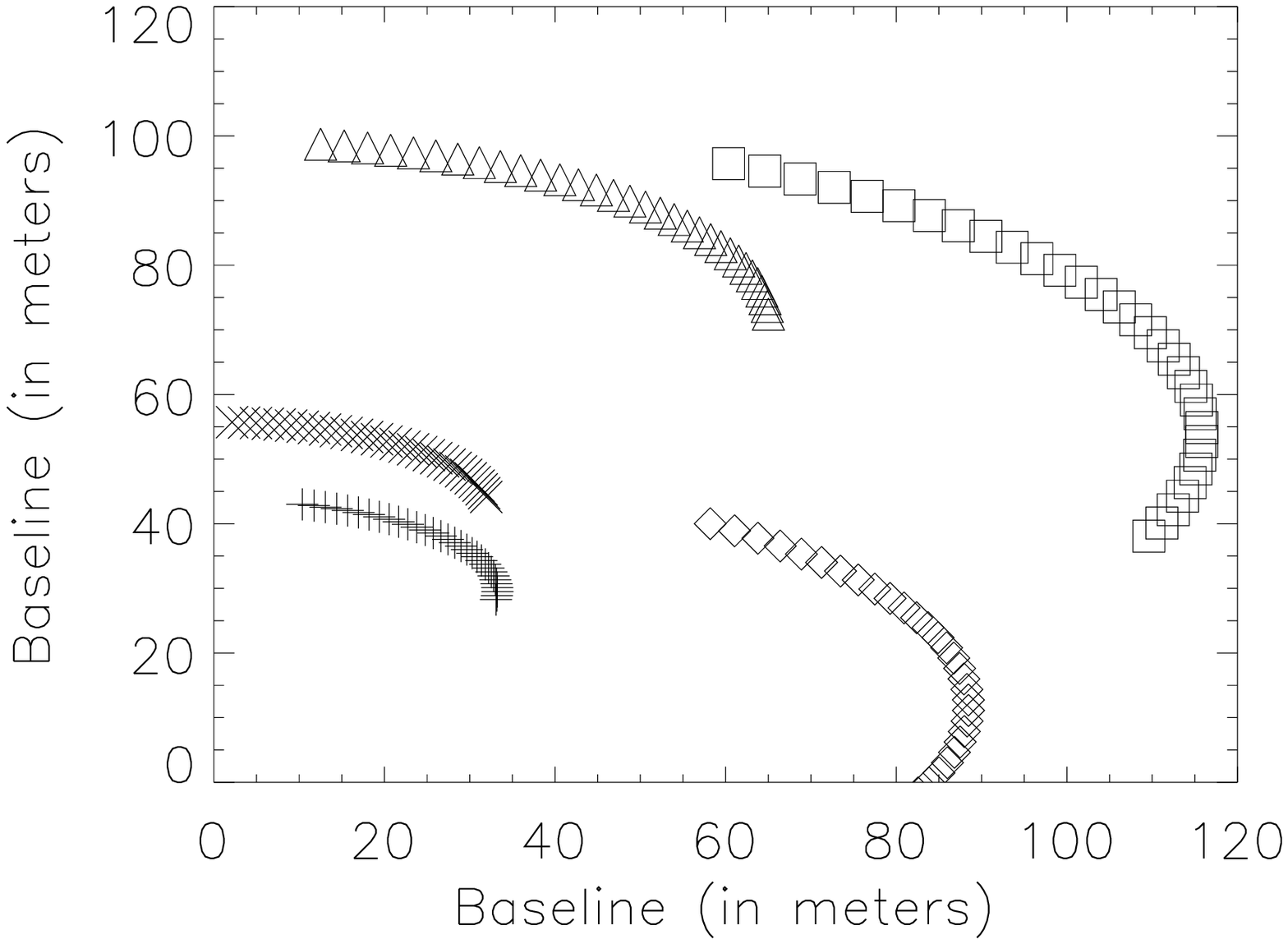}\\\includegraphics[width=8cm]{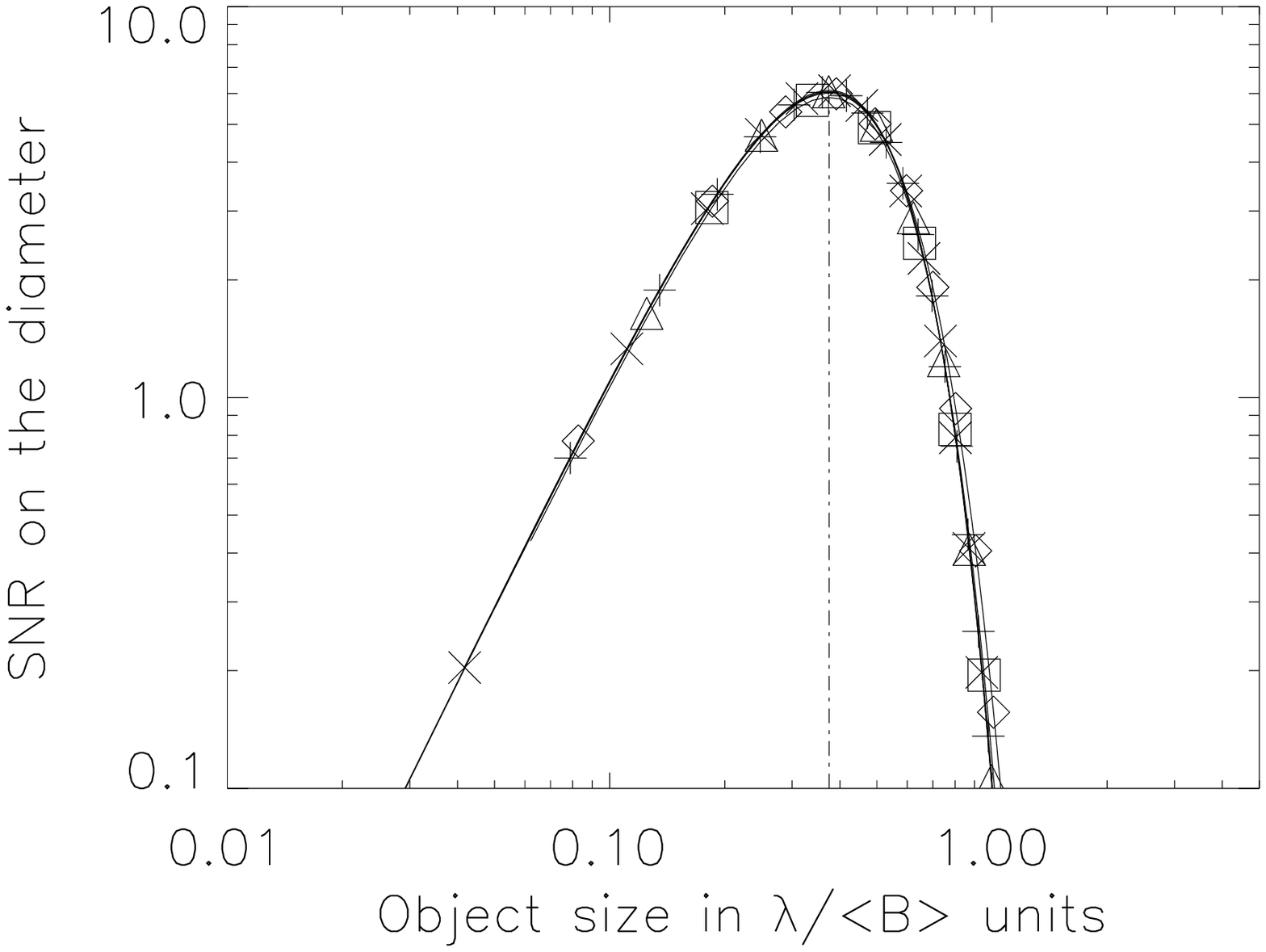}\\\includegraphics[width=8cm]{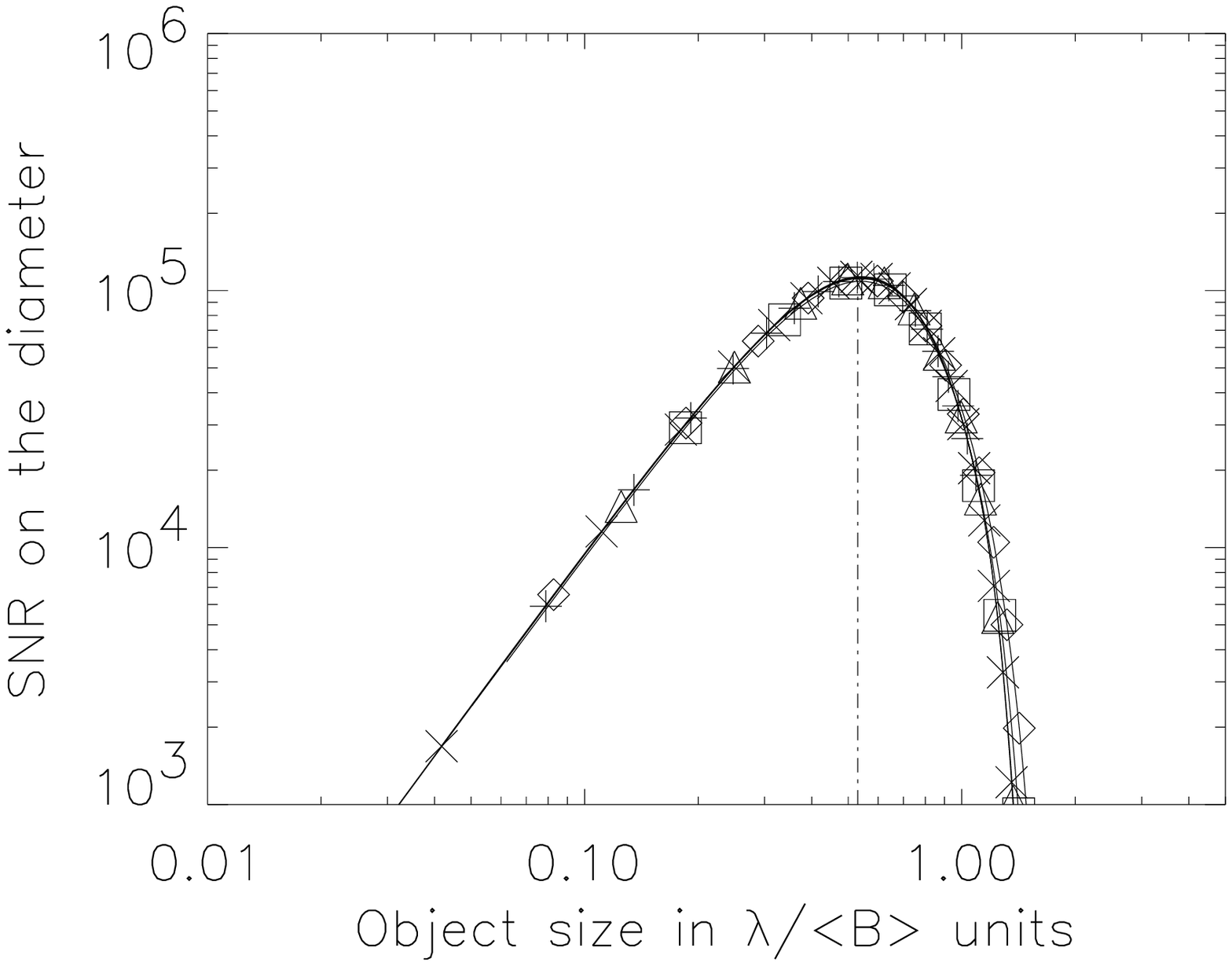}
 \end{tabular}
 \caption{\label{fig_snr_max_diam}Up: $(u,v)$ coverage
     for different UTs configurations. The average projected
baseline length
     is respectively: $B=45\mathrm{m}$ ($+$),  $B=56\mathrm{m}$
     ($\times$), $B=83\mathrm{m}$ ($\diamond$), $B=100\mathrm{m}$
     ($\triangle$), $B=124\mathrm{m}$ ($\square$). More informations
     about the different configurations considered are given in Table \ref{tab_uv}.
     Middle: 
     SNR of the diameter
     as a function of the size given in fraction of the 
     interferometer resolution $\lambda/B$, in the ``detector noise"
     regime.  Vertical line shows the size where the SNR is
     maximum. Bottom: same as above, but in the ``photon noise" regime.}
 \end{center}
 \end{figure}

We simulate different configurations with two telescopes that span the
range of average projected
baseline, respectively: (a) $B=45\mathrm{m}$ ($+$);
(b) $B=56\mathrm{m}$ ($\times$); (c) $B=83\mathrm{m}$ ($\diamond$);
(d) $B=100\mathrm{m}$ ($\triangle$); (e) $B=124\mathrm{m}$
($\square$). The declination of the source is arbitrarily set to
$-25^{\circ}$. The source is supposed to be observed between $-3\mathrm{h}$ to
$3\mathrm{h}$ from the zenith. The parameters of the different
configurations used are summarized in Table \ref{tab_uv} and corresponding $(u,v)$ plane coverages are shown in
Fig. \ref{fig_snr_max_diam}. (up).


We compute the SNR of the diameter for all these
configurations, in both ``detector noise" ($K=13$) and ``photon
noise" ($K=2$) regimes, respectively. In the ``detector noise" regime, 
the error of the visibility is independent on the diameter, and since
in this case Eq. \ref{eq_error_param} tells us that $\sigma(\widetilde{\theta})$ is directly inversely proportional
to $\ds \frac{\partial \mathcal{M}_{\theta}}{\partial \theta} \bigg|_{\theta=\widetilde{\theta}}$, 
the SNR of the diameter
is maximum when the product of the diameter by the 
derivative of the model is maximum too, namely when:
\begin{equation}
\frac{\partial}{\partial \theta}\left[\theta \frac{\partial \mathcal{M}_{\theta}}{\partial \theta}\right]
\Bigg|_{\theta=\widetilde{\theta}} = \frac{\partial}{\partial
  \theta}\left[\theta \frac{\partial |\widehat{O}_{\theta}|^2}{\partial \theta}\right]
\Bigg|_{\theta=\widetilde{\theta}}  = 0
\end{equation} 
This leads to:
\begin{equation}
\widetilde{\sigma}_o = \frac{2\sqrt{ln(2)}}{\pi\sqrt{2}}
\frac{\lambda}{<B>} \simeq 0.38 \frac{\lambda}{<B>} \label{eq_size_det}
\end{equation} 
where $<B>$ is th average projected baseline.

In the ``photon noise" regime, the 
error of the visibility is strongly dependent on the size. 
In that case, the maximum of
the SNR occurs when the projected baseline range scanned by the interferometer matches the 
object frequency content. Considering the frequency for which the visibility is $1/\mathrm{e}$, 
it comes:
\begin{equation}
\widetilde{\sigma}_o = \frac{2\sqrt{ln(2)}}{\pi}
\frac{\lambda}{<B>} \simeq 0.54 \frac{\lambda}{<B>} \label{eq_size_photon}
\end{equation} 
Fig. \ref{fig_snr_max_diam}. (middle and bottom) illustrates these results. The SNR
of the diameter is plotted as a function of the size in
$\lambda/B$ units, for the five geometrical configurations selected
above. We can see that the
whole curves present the same behavior, and especially the same
maximum. This maximum verifies Eq.'s \ref{eq_size_det} and \ref{eq_size_photon} for the ``detector noise" and
the ``photon noise" regimes, respectively. 
We conclude that the mean projected baseline 
optimizing the observation of an
object of diameter $\sigma_o$ is given by:

\begin{equation}
<B> \simeq 0.5 \frac{\lambda}{\sigma_o}
\end{equation}

Furthermore we stress that the performances degrades rapidly when this criterion is not respected.

\subsection{Effect of the correlation coefficients on the error bars} \label{sec_correlations}
We have shown in Section \ref{sec_noise} that the correlations between
the visibilities could reach a maximum value of $\rho = 0.5$. Since the estimator arising 
from maximum likelihood (i.e. the $\chi^2$
minimization) is unbiased, the expected value of the fitted parameters
does not depend on whether the correlation coefficients are introduced or not.
We analyze here the effect of the correlation coefficients on the derived error.

\begin{figure}[!*p]
\begin{center}
\includegraphics[width=12cm]{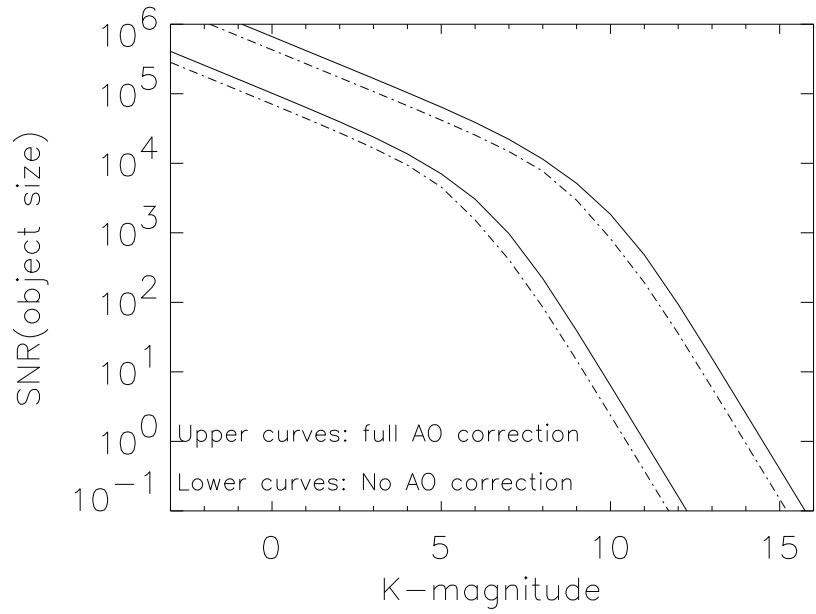}
\caption{\label{fig_w_wo_covariance}SNR of the diameter
    as a 
    function of the magnitude, for perfect AO correction
    (upper curves, $\sigma_{\mathcal{S}}=0$) and in the pure turbulent
    case (lower curves,
    $\sigma_{\mathcal{S}}=\overline{\mathcal{S}}$). SNR is computed
    assuming diagonal covariance of the measurements (no correlation:
    solid line) or considering correlation coefficients as computed in
    Section \ref{sec_noise} (dashed lines).}
\end{center}
\end{figure}

We simulate the observation of a Gaussian source with three
UTs, together with the instrumental parameters of Section
\ref{sec_applications}
and we 
choose $\sigma_o = 0.5\lambda/<B>$ for the FWHM of the source.
Fig. \ref{fig_w_wo_covariance}. shows, in both pure
turbulent ($\sigma_{\mathcal{S}}=\overline{\mathcal{S}}$) and fully AO
  corrected ($\sigma_{\mathcal{S}}=0$) cases, the SNR of $\sigma_o$,
with and without taking into account the correlation coefficients. 
It results that considering uncorrelated measurements leads to
underestimate the error (or overestimate the SNR) by a factor of
$\sqrt{2}$, roughly. Given that the ratio of the null elements 
versus the non null elements in the covariance matrix increases with the number of telescopes,
we infer that this factor $\sqrt{2}$ is an upper limit. 

\section{Summary}
We have computed 
the theoretical covariance matrices of the modal visibility and the
modal closure phase in the presence of partial AO correction, 
when the measurements
are corrupted by atmospheric, photon
and detector noises. In the photon noise regime and for an interferometer with a large number of telescopes,
the measurements are most of the time uncorrelated. In any case, the correlation coefficients are always 
smaller than $1/2$ for the visibilities and smaller than $1/3$ for the closure phases.

Then from a classical least square approach,
we have investigated the ability of interferometers to measure stellar diameters.
In the light of the AMBER experiment we have
found a limiting magnitude in the range $K=11-13$ depending on the size of the source. At last, we have shown  
that the observation is optimized when the mean resolution of the interferometer 
is equal to twice the stellar diameter.

\renewcommand{\theequation}{A-\arabic{equation}}
 \setcounter{equation}{0}  
\appendix
\section*{APPENDIX A: STATISTICS OF THE OBSERVABLES} \label{app_noise}
\subsection*{1. General formalism}
In order to compute the moments of the spectral density, we use the
spatially continuous model of photodetection process of Goodman\cite{goodman_1} where the detected signal is corrupted by photon
noise, by additive Gaussian noise $\epsilon$ of variance $\sigma^2$
and by  the turbulent
atmosphere. It takes the form:
\begin{equation}
s(x,y) = \sum_{n=1}^{K}\delta(x-x_n) + \epsilon(x)
\end{equation}
and its Fourier transform: 
\begin{equation}
 \widehat{S}(f) = \sum_{n=1}^{K}\exp\left(-2i\pi{fx_n}\right) +  \widehat{\epsilon}(f)\label{eq_dfourier}
\end{equation}
\subsection*{2. The modal visibility}
Such calculation has been already done in Tatulli et
al\cite{tatulli_1} . We give
here the results assuming that the low and high coupling coefficients
verify Eq. \ref{eq_rhoi_simplified} and \ref{eq_rhoij_simplified}
respectively. We remind that following equations assume also that the telescope
transmissions are all equal, i.e. $K_i = K/N_{tel}$, where $K_i$ is
the number of photoevents coming from telescope $i$ and $N_{tel}$ is
the number of the telescope, and that the level of corrections of the
Adaptive Optics systems are the same for all the telescopes. 

The square relative error of the modal
visibility can be seen as the sum of two contributions:
\begin{equation}
\frac{\sigma^2\{V_{ij}^2\}}{\overline{V_{ij}^2}^2} = \mathcal{E}^2_P(K,\mathcal{S}) 
+ \mathcal{E}^2_A(K, \sigma^2, \mathcal{S})
\end{equation}
with $\mathcal{E}^2_P$ the photon noise square relative error and
$\mathcal{E}^2_A$ the additive noise   relative error, as described in
Tatulli et al\cite{tatulli_1} .
The same way, the covariance can be cut in two terms:
\begin{equation}
\frac{\mathrm{Cov}\left\{V_{ij}^2,V_{kl}^2\right\}}{\overline{V_{ij}^2}\overline{V_{kl}^2}} = \mathcal{C}_P(K,\mathcal{S}) 
+ \mathcal{C}_A(K, \sigma^2, \mathcal{S})
\end{equation}
Here we must pay attention that two cases can occur (see Fig.
\ref{fig_vis_config}. in the body of the text): (i) either the
baselines $f_{ij}$ and $f_{kl}$ come from two distinct pairs of telescopes
(ii) either the baselines have one telescope in common, let say $j=k$,
which drives to extra correlation between the visibilities.

The results, i.e. the diagonal and non diagonal terms of the visibility
covariance matrix, are summarized in table \ref{table_covis}.

\subsection*{3. The closure phase}
The estimator of the bispectrum between telescopes $i,j,k$ is defined
by:
\begin{equation}
Q_{ijk}= \widehat{S}(f_{ij}) \widehat{S}(f_{jk}) \widehat{S}^{\ast}(f_{ik})
\end{equation}

Chelli\cite{chelli_1} has shown that the covariance (per sample) on the
closure phase does only depend on the  modulus of the bispectrum, 
and hence assuming a
centro-symmetrical source,  could be written:
\begin{equation}
\mathrm{Cov}\left\{\Phi_{ijk},\Phi_{lmn}\right\} =\frac{1}{2} \frac{\mathrm{Re}[Q_{ijk}(Q_{lmn}-Q^{\ast}_{lmn})]}{\mathrm{E}(Q_{ijk})\mathrm{E}(Q_{lmn})}
\end{equation}
with 
\begin{eqnarray}
\mathrm{E}(Q_{ijk}) &=& <\overline{K}^3\widehat{i}(f_{ij})\widehat{i}(f_{jk})\widehat{i}^{\ast}(f_{kl})>_{\phi}\\
&=& \overline{\mathcal{S}}^3[V_{\star}(f_{ij})V_{\star}(f_{jk})V_{\star}^{\ast}(f_{ik})]
\frac{\rho_0^3\overline{K}^3}{N_{tel}^3}
\end{eqnarray}
where $\widehat{i}(f)$ is the normalized spectral density.

For sake of simplicity we analyze the  second order
moments of the estimator in two cases separately: the ``photon noise"
case and the ``detector noise" case. Then the statistics with regards to
the atmosphere are taken into account.

\subsection*{4. The photon noise case}
Following Goodman's formalism\cite{goodman_1}$^,$\cite{chelli_1} the
second order moments are:
\begin{eqnarray}
\mtotK\{|Q_{ijk}|^2\} &=& \overline{K}^3 + \overline{K}^4\left[|\widehat{i}(f_{ij})|^2+|\widehat{i}(f_{jk})|^2+|\widehat{i}(f_{ik})|^2\right] \nonumber \\
 && + \overline{K}^5 \left[|\widehat{i}(f_{ij})|^2|\widehat{i}(f_{jk})|^2 +
 |\widehat{i}(f_{ij})|^2|\widehat{i}(f_{ik})|^2+|\widehat{i}(f_{jk})|^2|\widehat{i}(f_{ik})|^2\right] \nonumber \\
 &&+ \overline{K}^6\left[|\widehat{i}(f_{ij})|^2|\widehat{i}(f_{jk})|^2|\widehat{i}(f_{ik})|^2\right]
 \end{eqnarray}
 \begin{eqnarray}
   \mtotK\{Q_{12}^2\} &=&2\overline{K}^3\left[\widehat{i}^{\ast}(f_{ij})\widehat{i}^{\ast}(f_{jk})\widehat{i}(f_{ik})\right] + \overline{K}^4\left[|\widehat{i}(f_{ij})|^4+|\widehat{i}(f_{jk})|^4+|\widehat{i}(f_{ik})|^4\right. \nonumber \\ 
 &&+\left. 2|\widehat{i}(f_{ij})|^2|\widehat{i}(f_{jk})|^2+2|\widehat{i}(f_{ij})|^2|\widehat{i}(f_{ik})|^2+2|\widehat{i}(f_{jk})|^2|\widehat{i}(f_{ik})|^2\right] \nonumber \\
 && + 2\overline{K}^5\left[\widehat{i}(f_{ij})\widehat{i}(f_{jk})\widehat{i}^{\ast}(f_{ik})|\widehat{i}(f_{ij})|^2+|\widehat{i}(f_{jk})|^2+|\widehat{i}(f_{ik})|^2)\right]  \nonumber \\
 && +\overline{K}^6\left[\widehat{i}(f_{ij})\widehat{i}(f_{jk})\widehat{i}^{\ast}(f_{ik})\right]^2
 \end{eqnarray}

For the covariance, two cases may appear (see Fig. \ref{fig_cov_config}.
in the body of the text): (i) both triplets of
telescopes ($i,j,k$ and $l,m,n$) are different or 
(ii) two telescopes are part of both
triplets, in that case there is a baseline in common, let say
$f_{ij}=f_{lm}$. This add extra terms in the covariance.

\begin{eqnarray}
\mtotK\{Q_{ijk}Q_{lmn}^{\ast}\}
&=&\overline{K}^6\left[\widehat{i}(f_{ij})\widehat{i}(f_{jk})\widehat{i}^{\ast}(f_{ik})\widehat{i}^{\ast}(f_{lm})\widehat{i}^{\ast}(f_{mn})\widehat{i}(f_{ln})\right] \nonumber \\
&&{\mathrm{if}~~f_{ij}=f_{lm}} \nonumber \\
&& + \overline{K}^3|\widehat{i}(f_{kn})|^2 +
\overline{K}^4|\widehat{i}(f_{ij})|^2|\widehat{i}(f_{kn})|^2 \nonumber \\
&&
 + \overline{K}^4\left[\widehat{i}(f_{kn})\widehat{i}(f_{ik})\widehat{i}^{\ast}(f_{ln})+
  \widehat{i}(f_{kn})\widehat{i}(f_{jk})\widehat{i}^{\ast}(f_{mn})\right]\nonumber
\\
&& + \overline{K}^5\left[\widehat{i}(f_{jk})\widehat{i}^{\ast}(f_{ik})\widehat{i}^{\ast}(f_{mn})\widehat{i}(f_{ln})\right]\nonumber \\
&& + \overline{K}^5|\widehat{i}(f_{ij})|^2\left[\widehat{i}(f_{kn})\widehat{i}(f_{ik})\widehat{i}^{\ast}(f_{ln})+\widehat{i}(f_{kn})\widehat{i}(f_{jk})\widehat{i}^{\ast}(f_{mn})\right]
 \end{eqnarray}
\begin{eqnarray}
\mtotK\{Q_{ijk}Q_{lmn}\}
&=&\overline{K}^6\left[\widehat{i}(f_{ij})\widehat{i}(f_{jk})\widehat{i}^{\ast}(f_{ik})\widehat{i}(f_{lm})\widehat{i}(f_{mn})\widehat{i}^{\ast}(f_{ln})\right]\nonumber  \\
&&{\mathrm{if}~~f_{ij}=f_{lm}} \nonumber \\
&&  +
\overline{K}^4\left[|\widehat{i}(f_{mn})|^2|\widehat{i}(f_{ik})|^2 + |\widehat{i}(f_{mn})|^2|\widehat{i}(f_{jk})|^2\right] \nonumber \\
&&
 + \overline{K}^4\left[|\widehat{i}(f_{ln})|^2|\widehat{i}(f_{ik})|^2 + |\widehat{i}(f_{ln})|^2|\widehat{i}(f_{jk})|^2\right]\nonumber
\\
&& + \overline{K}^5\left[\widehat{i}(f_{jk})\widehat{i}^{\ast}(f_{ik})\widehat{i}^{\ast}(f_{mn})\widehat{i}(f_{ln})\right]\nonumber \\
&& +
\overline{K}^5\left[\widehat{i}(f_{ij})\widehat{i}(f_{jk})\widehat{i}^{\ast}(f_{ik})(|\widehat{i}(f_{mn})|^2+|\widehat{i}(f_{ln})|^2)\right]\nonumber \\
&& + \overline{K}^5\left[\widehat{i}(f_{lm})\widehat{i}(f_{mn})\widehat{i}^{\ast}(f_{ln})(|\widehat{i}(f_{jk})|^2+|\widehat{i}(f_{ik})|^2)\right]
 \end{eqnarray}

\subsection*{5. The detector noise case}
We suppose a zero mean Gaussian detector noise of variance $\sigma^2$.
\begin{eqnarray}
\meps\{|Q_{ijk}|^2\} &=&
 \overline{K}^6\left[|\widehat{i}(f_{ij})|^2|\widehat{i}(f_{jk})|^2|\widehat{i}(f_{ik})|^2\right]
  \nonumber\\
&& + N\sigma^2 \overline{K}^4\left[|\widehat{i}(f_{ij})|^2|\widehat{i}(f_{jk})|^2 +
 |\widehat{i}(f_{ij})|^2|\widehat{i}(f_{ik})|^2+|\widehat{i}(f_{jk})|^2|\widehat{i}(f_{ik})|^2\right]
 \nonumber\\
&& +
 (3N\sigma^4+N^2\sigma^4)\overline{K}^2\left[|\widehat{i}(f_{ij})|^2|\widehat{i}(f_{jk})|^2|\widehat{i}(f_{ik})|^2\right]  \nonumber\\
&& + N^3\sigma^6+3N^2\sigma^6+15N\sigma^6
\end{eqnarray}
\begin{eqnarray}
 \meps\{Q_{ijk}^2\} &=&
 \overline{K}^6\left[\widehat{i}(f_{ij})\widehat{i}(f_{jk})\widehat{i}^{\ast}(f_{ik})\right]^2+15N\sigma^6
\end{eqnarray}
\begin{eqnarray}
\meps\{Q_{ijk}Q_{lmn}^{\ast}\}
&=&
\overline{K}^6\left[\widehat{i}(f_{ij})\widehat{i}(f_{jk})\widehat{i}^{\ast}(f_{ik})\widehat{i}^{\ast}(f_{lm})\widehat{i}^{\ast}(f_{mn})\widehat{i}(f_{ln})\right]
 + 15N\sigma^6\nonumber \\
&&{\mathrm{if}~~f_{ij}=f_{lm}} \nonumber \\
&& + N\sigma^2
\overline{K}^4\left[\widehat{i}(f_{ij})\widehat{i}^{\ast}(f_{ik})\widehat{i}^{\ast}(f_{mn})\widehat{i}(f_{ln})\right]
+ 3N^2\sigma^6
\end{eqnarray}
\begin{eqnarray}
\meps\{Q_{ijk}Q_{lmn}\} &=& \overline{K}^6\left[\widehat{i}(f_{ij})\widehat{i}(f_{jk})\widehat{i}^{\ast}(f_{ik})\widehat{i}(f_{lm})\widehat{i}(f_{mn})\widehat{i}^{\ast}(f_{ln})\right]
+ 15N\sigma^6
\end{eqnarray}

Once taking into account the statistics with regards to the
atmospheric turbulence, the final expression are given in both ``photon
noise" and ``detector noise" regime in tables \ref{table_covclotphot} and \ref{table_covclotdet}, respectively.

The expression of the covariance in the general case is obtained by
adding the covariance in each regime:
\begin{equation}
\mathrm{Cov}\left\{\Phi_{ijk},\Phi_{lmn}\right\} =
\mathrm{Cov}_{phot}\left\{\Phi_{ijk},\Phi_{lmn}\right\}  + \mathrm{Cov}_{det}\left\{\Phi_{ijk},\Phi_{lmn}\right\} 
\end{equation}
\renewcommand{\theequation}{B-\arabic{equation}}
 \setcounter{equation}{0}  
\section*{APPENDIX B: MODELLING PARTIAL ADAPTIVE OPTICS CORRECTION}\label{app_strehl}
We consider a simplified model where the
long exposure AO corrected transfer function can be divided in two
weighted components: one perfect transfer function and one turbulent transfer
function where the weight $h$ describes the strength of the correction: 
\begin{eqnarray}
<T^i(f)> &=&  \frac{T^i_0(f)}{\int T^i_0(f)\du{f}}[h + (1-h)\mathcal{B}_{\phi}(f)] ,~h \in [0,1]\nonumber \\
 &=&  \frac{T^i_0(f)}{\int T^i_0(f)\du{f}} \exp\left[-\frac{1}{2}\mathcal{D}_{\phi}(f)\right]
\end{eqnarray}
where $\mathcal{D}_{\phi}(f)$ is the AO corrected structure function.
For sake of simplicity we assume that the transfer
function of the turbulence $\mathcal{B}_{\phi}(f)$ is Gaussian, hence:
\begin{equation}
\mathcal{B}_{\Phi} =
\exp\left(-\frac{f^2}{\sigma_{\mathcal{B}}^2}\right),
\sigma_{\mathcal{B}}=\sqrt{\frac{2}{6.88}}\frac{r_0}{\lambda} \label{eq_bfunc}
\end{equation}
Going further we can notice that the AO corrected transfer function of
the atmosphere can be decomposed in a low frequency term (halo) and a high
frequency term (coherent energy)\cite{conan_1} :
\begin{equation}
\exp\left[-\frac{1}{2}\mathcal{D}_{\phi}(f)\right] =
\mathrm{FTO}_{halo} + \exp(-\sigma^2_{\phi}) 
\end{equation}
and that,
\begin{equation}
\exp\left[-\frac{1}{2}\mathcal{D}_{\phi}(f)\right] = h +
(1-h)\mathcal{B}_{\phi}(f) \label{eq_struct_fonc}
\end{equation} 
Hence we have:
\begin{equation}
h = \exp(-\sigma^2_{\phi}) = \mathrm{E}_c(D/r_0, N_{z})
\end{equation}
where the coherent energy $\mathrm{E}_c$ depends on  the level of
correction (number of Zernike $N_{z}$) and the strength of the
turbulence $D/r_0$.

\begin{figure}[t]
\begin{center}
\begin{tabular}{c}
\includegraphics[width=10cm]{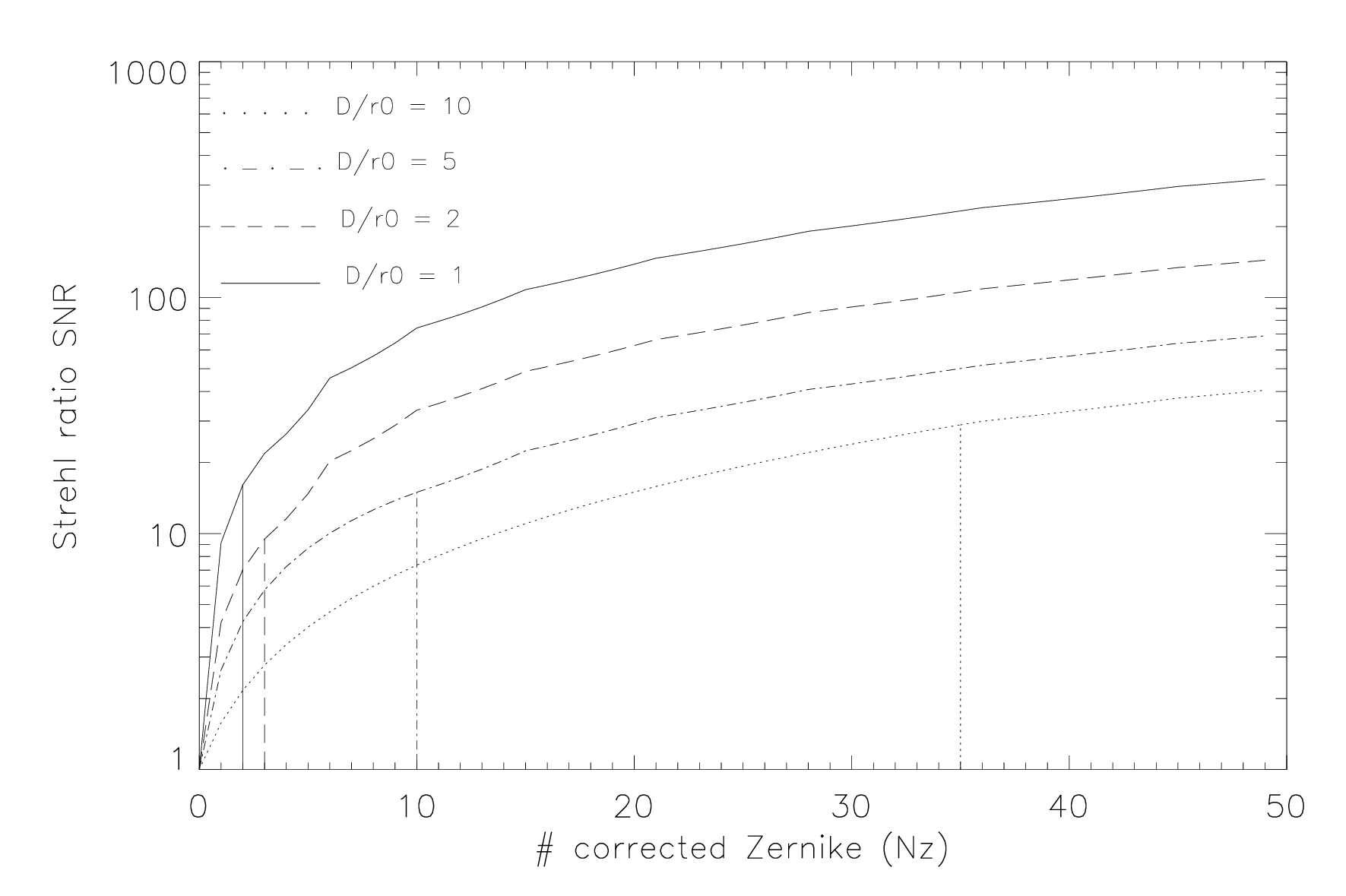} \\ \includegraphics[width=10cm]{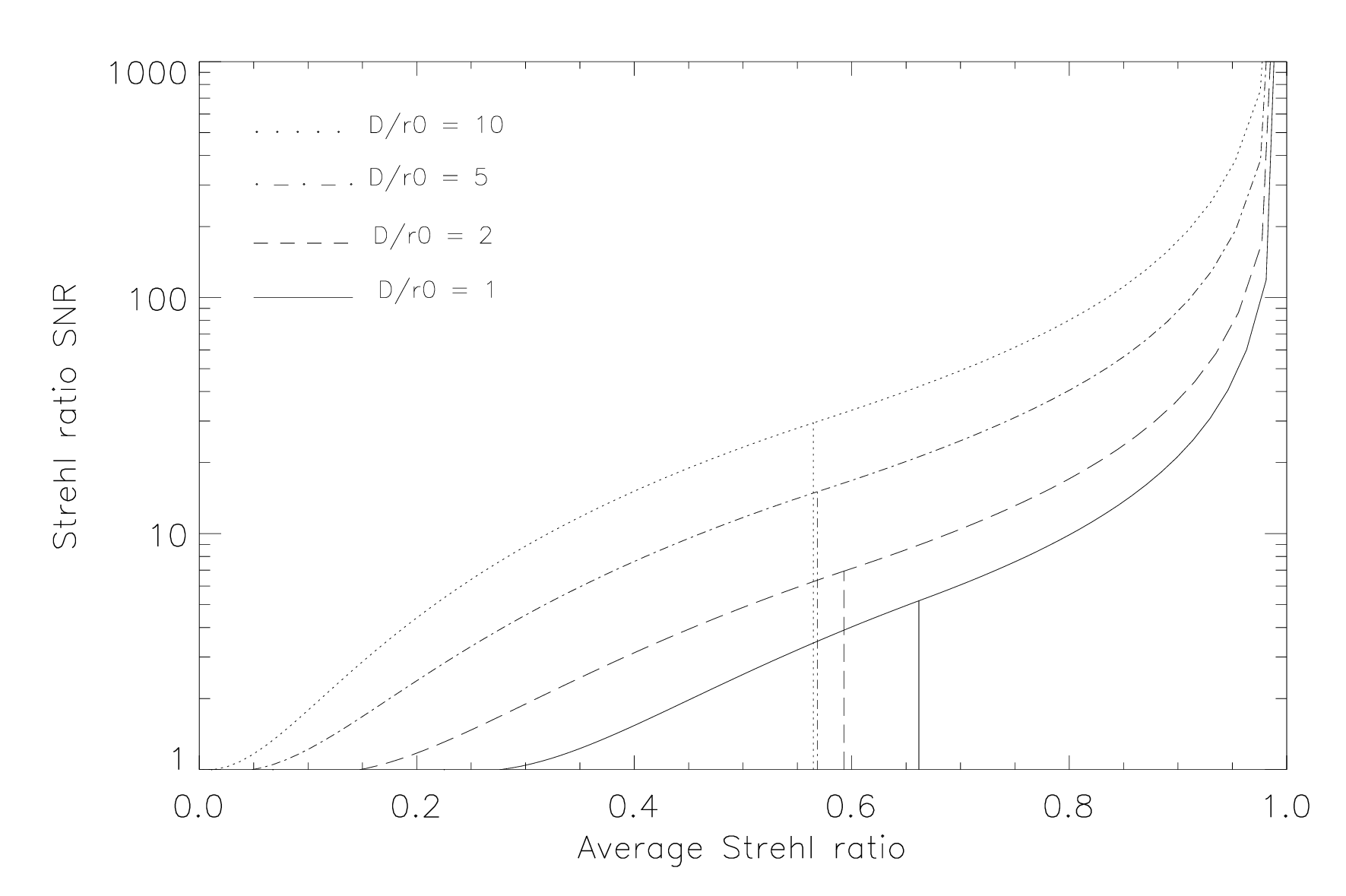}
\end{tabular}
\caption{\label{fig_strehl_stat}SNR of the Strehl
    ratio  as a function  of the level of correction in number of
    corrected Zernike (top), and as a function of the long exposure
    Strehl ratio (bottom). Results are given for $D/r_0=1$ (solid line),
    $D/r_0=2$ (dashed line), $D/r_0=5$ (dash-dotted line), $D/r_0=10$
    (dotted line). Vertical lines show from where the interpolation
    has been performed.}
\end{center}
\end{figure}

Using the definition of the long exposure Strehl ratio:
\begin{equation}
\overline{\mathcal{S}} = \int <T^i(f)> \du{f}
\end{equation}
 it comes:
\begin{equation}
\overline{\mathcal{S}} = h + (1-h)
\frac{\sigma^2_{\mathcal{B}}}{\sigma^2_{\mathcal{B}}+\sigma^2_{T}} = \mathrm{E}_c
+
\frac{(1-\mathrm{E}_c)}{1+3.44\left(\frac{D}{r_0}\right)^2}
\end{equation}
The second order moment of the instantaneous Strehl ratio writes:
\begin{eqnarray}
\overline{\mathcal{S}^2} &=& \iint <T^i(f)T^i(f^{'})> \du{f}\du{f^{'}} 
 \nonumber \\
&=&\iint P_i(r)P_i(r+\rho)P_i(r^{'})P_i(r^{'}+\rho^{'}) \nonumber \\
&& \times<\Psi_i(r)\Psi_i^{\ast}(r+\rho)\Psi_i^{\ast}(r^{'})\Psi_i(r^{'}+\rho^{'})> \du{r}\du{\rho}\du{r^{'}}\du{\rho^{'}} 
 \end{eqnarray}
where  $P_i(r)$ is the pupil function and $\Psi_i(r)$ is the complex amplitude of
the AO corrected wavefront. We use Korff's\cite{korff_1} derivation of the moments of the complex
amplitude of the wavefronts to introduce in the
equations linear combinations of the structure function
($\mathcal{D}(r)$) at different
spatial frequencies. Assuming the structure function to be
stationary, it comes:
\begin{eqnarray}
\overline{\mathcal{S}^2} &=&
P_i(r)P_i(r+\rho)P_i(r^{'})P_i(r^{'}+\rho^{'}).\nonumber \\
&&
\frac{\exp\left\{-\frac{1}{2}[\mathcal{D}(\rho)+\mathcal{D}(\rho^{'})+\mathcal{D}(r^{'}-r)+\mathcal{D}(r^{'}+\rho^{'}-r-\rho)]\right\}}{\exp\left\{-\frac{1}{2}[\mathcal{D}(r^{'}+\rho^{'}-r)+\mathcal{D}(r^{'}-r-\rho)]\right\}}
\du{r}\du{\rho}\du{r^{'}}\du{\rho^{'}} 
\end{eqnarray}
Using Eq. \ref{eq_struct_fonc}, previous expression can be
analytically developed for $h=\mathrm{E_c}=]0.5,1]$, i.e. for good AO
correction levels\cite{tatulli_1} . 
However we know that, without AO correction  and
assuming circular Gaussian statistics for the complex amplitude of the
wavefront, we have $\sigma_{\mathcal{S}} =
\overline{\mathcal{S}}$\cite{roddier_1}$^,$\cite{goodman_1} .
Hence we can perform an interpolation between the good AO corrections and
the pure turbulent cases. Fig \ref{fig_strehl_stat}. shows the SNR of the Strehl ratio
as a function respectively of the number of Zernike (i.e. the
correction level) and the average Strehl ratio $\overline{\mathcal{S}}$, at different
$D/r_0$. For each curve, we point out where from the interpolation has
been done. 

\clearpage
  \begin{table*}[!*p]
\begin{center}
\caption{\label{table_error}Variance of the visibility and
    the closure phase for a point source. 
$N_{pix}$ corresponds to the number of pixels
    that sample the interferogram. $N_{pix} \ge 2$ is required to
    respect the Shannon
    criterion. $\sigma$ is the detector noise per pixel.}
\begin{tabular}{|c|c|c|c|c|} \hline
\multicolumn{2}{|c|}{}  &  \multicolumn{3}{|c|}{}\\
\multicolumn{2}{|c|}{} &  \multicolumn{3}{|c|}{Variance of the
  observables (point source)}  \\ 
\multicolumn{2}{|c|}{}  &  \multicolumn{3}{|c|}{} \\\cline{3-5} 
\multicolumn{2}{|c|}{}  &  \multicolumn{2}{|c|}{} &\\  
\multicolumn{2}{|c|}{Observables} &  \multicolumn{2}{|c|}{Photon noise regime ($\overline{K} \gg 1$)}&  \\ 
\multicolumn{2}{|c|}{}  &  \multicolumn{2}{|c|}{} & Detector noise regime
\\ \cline{3-4} 
\multicolumn{2}{|c|}{}  &   \multicolumn{1}{|c|}{} & &($\overline{K}
\ll 1$) \\
\multicolumn{2}{|c|}{}  &
Full AO correction&
No AO correction & \\
\multicolumn{2}{|c|}{}  &   \multicolumn{1}{|c|}{} & & \\\hline \multicolumn{2}{|c|}{}&&&\\
\multicolumn{2}{|c|}{$\ \widetilde{V^2}$}  & $
\left[\frac{2N_{tel}}{(1-\tau)}+\frac{2}{\tau}\right]\frac{N_{tel}}{\rho_0\overline{\mathcal{S}}\overline{K}}$
&
$ \left[\frac{2N_{tel}+4}{(1-\tau)}+\frac{4}{\tau}\right]\frac{N_{tel}}{\rho_0\overline{\mathcal{S}}\overline{K}}$
&   $ \left[\frac{3N_{pix}\sigma^4 + N_{pix}^2\sigma^4}{(1-\tau)^4}\right]\frac{N_{tel}^4}{\rho_0^4\overline{\mathcal{S}}^4\overline{K}^4}$\\  \multicolumn{2}{|c|}{}&&&\\
\multicolumn{2}{|c|}{$ \widetilde{\phi}_{B}$} &  $
\left[\frac{3N_{tel}-6}{2}\right]\frac{N_{tel}}{\rho_0\overline{\mathcal{S}}\overline{K}}$
& $
\left[\frac{3N_{tel}-3}{2}\right]\frac{N_{tel}}{\rho_0\overline{\mathcal{S}}\overline{K}}$
&  $ \left[\frac{9N_{pix}\sigma^4 + 3N_{pix}^2\sigma^4}{2}\right]\frac{N_{tel}^4}{\rho_0^4\overline{\mathcal{S}}^4\overline{K}^4}$ \\\multicolumn{2}{|c|}{}&&&\\
\hline
 \end{tabular}
\end{center}
\end{table*}

\clearpage

\begin{table*}[!p]
\begin{center}
\caption{\label{table_noises}Correlation coefficients of the visibilities and
    the closure phases. For the visibility, two cases are considered:
    one telescope is common to both baselines, hence the triplet of
    telescopes is forming a triangle (a so-called closure), or not
    (see Fig. \ref{fig_vis_config}.). For the closure phases, two cases are
    investigated as well: one baseline belongs to both closure
    phases or not (see Fig. \ref{fig_cov_config}).}
\begin{tabular}{|c|c|c|c|c|} \hline
\multicolumn{2}{|c|}{}  &  \multicolumn{3}{|c|}{}\\
\multicolumn{2}{|c|}{} &  \multicolumn{3}{|c|}{Correlation coefficient
  $\rho$ (point source)}  \\ 
\multicolumn{2}{|c|}{}  &  \multicolumn{3}{|c|}{} \\\cline{3-5} 
\multicolumn{2}{|c|}{}  &  \multicolumn{2}{|c|}{} &\\  
\multicolumn{2}{|c|}{Observables} &  \multicolumn{2}{|c|}{Photon noise regime ($\overline{K} \gg 1$)}&  \\ 
\multicolumn{2}{|c|}{}  &  \multicolumn{2}{|c|}{} & Detector noise regime
\\ \cline{3-4} 
\multicolumn{2}{|c|}{}  &   \multicolumn{1}{|c|}{} & &($\overline{K}
\ll 1$) \\
\multicolumn{2}{|c|}{}  &
Full AO correction &
No AO correction & \\
\multicolumn{2}{|c|}{}  &   \multicolumn{1}{|c|}{} & & \\\hline
 &  if closure  & $\ds \leqslant \frac{1}{2}
 $  & $\ds \leqslant  \frac{1}{4}
 $  & \\
$\ds \widetilde{V^2}$  & &  &  & $\ds \frac{3}{3+N_{pix}}$\\ 
 &otherwise &   $\ds 0$  & $\ds 0$ &\\\hline
&  if baseline in common &$\ds \frac{1}{3}$&
 $\ds \leqslant \frac{1}{6}$  &  $\ds \frac{3}{9+N_{pix}}$ \\
$\ds \widetilde{\phi}_{B}$ & &  & & \\
&  otherwise &  $\ds 0$ & $\ds 0$ & $\ds 0$\\\hline
 \end{tabular}
\end{center}
\end{table*}

\clearpage

\begin{table*}[!p]
\begin{center}
\caption{\label{tab_uv} Description of the considered telescopes 
configurations. Are given the name of the telescopes pairs, their
respective average projected baseline as well as the related symbols
used in Fig. \ref{fig_snr_max_diam}. The telescopes are the four $8$ meters Unit telescopes
(UTs) of the VLTI. The declination of the source is arbitrarily set to
$-25^{\circ}$. The source is supposed to be observed between $-3\mathrm{h}$ to
$3\mathrm{h}$ from the zenith.}  
\begin{tabular}{|c|c|c|}
\hline \hline
Symbol & Telescopes & Average projected baseline \\
\hline \hline
$+$ & UT2-UT3 & $45\mathrm{m}$\\
$\times$ &  UT1-UT2 & $56\mathrm{m}$ \\
$\diamond$ & UT2-UT4 & $83\mathrm{m}$\\
$\triangle$ & UT1-UT3 & $100\mathrm{m}$\\
$\square$ & UT1-UT4 & $124\mathrm{m}$ \\ \hline
\end{tabular}
\end{center}
\end{table*}

\clearpage

\begin{sidewaystable}
\centering
\caption{\label{table_covis}Description of the elements
    of the visibility covariance matrix.}
{\scriptsize
\begin{tabular}{|c|c|l|}
\hline
&  \multicolumn{2}{|c|}{}\\
Observables &  \multicolumn{2}{|c|}{Covariance coefficients}  \\
&  \multicolumn{2}{|c|}{}\\
\hline
&&\\
 & Diagonal  & $\ds \left[\frac{4\sigma^2_{\mathcal{S}}+2N_{tel}\overline{\mathcal{S}}^2}{(1-\tau)V_{\star}^2(f_{ij})\overline{\mathcal{S}}^3}
  +\frac{2(\sigma^2_{\mathcal{S}}+\overline{\mathcal{S}}^2)}{\tau\overline{\mathcal{S}}^3}\right]\frac{N_{tel}}{\rho_0\overline{K}}
+ 
\left[\frac{2N_{tel}\sigma^2_{\mathcal{S}}+N_{tel}^2\overline{\mathcal{S}}^2}{(1-\tau)^2V_{\star}^4(f_{ij})\overline{\mathcal{S}}^4}
  +\frac{4}{(1-\tau)^2V_{\star}^2(f_{ij})\overline{\mathcal{S}}^2}
  +\frac{1}{\tau^2\overline{\mathcal{S}}^2}\right]
\frac{N_{tel}^2}{{\rho_0^2\overline{K}^2}}
+
 \frac{1}{(1-\tau)^3V_{\star}^4(f_{ij})\overline{\mathcal{S}}^3}\frac{N_{tel}^4}{{\rho_0^3\overline{K}^3}}$\\
Visibility & coefficients& $\ds + 
 \left[\frac{2N\sigma^2}{(1-\tau)^2V_{\star}^2(f_{ij})\overline{\mathcal{S}}^2}\right]\frac{N_{tel}^2}{\rho_0^2\overline{K}^2}
+
 \left[\frac{2N\sigma^2}{(1-\tau)^3V_{\star}^4(f_{ij})\overline{\mathcal{S}}^3}\right]\frac{N_{tel}^4}{\rho_0^3\overline{K}^3}
 +  \left[\frac{3N\sigma^4 + N^2\sigma^4}{(1-\tau)^4V_{\star}^4(f_{ij})\overline{\mathcal{S}}^4}\right]\frac{N_{tel}^4}{\rho_0^4\overline{K}^4}$
\\
&&\\
\cline{2-3}
&&\\
$\ds \frac{\mathrm{Cov}\left\{V_{ij}^2,V_{kl}^2\right\}}{\overline{V_{ij}^2}\overline{V_{kl}^2}}$
& Non diagonal  &
$\ds
 \left[\frac{2}{(1-\tau)^2\overline{\mathcal{S}}^2}\left[\frac{1}{V_{\star}^2(f_{ij})}+\frac{1}{V_{\star}^2(f_{kl})}\right]
+
\frac{N_{tel}\sigma^2_{\mathcal{S}}}{(1-\tau)^2V_{\star}^2(f_{ij})V_{\star}^2(f_{kl})\overline{\mathcal{S}}^4}\right]\frac{N_{tel}^2}{\rho_0^2\overline{K}^2} 
+
 \left[\frac{1}{(1-\tau)^3V_{\star}^2(f_{ij})V_{\star}^2(f_{kl})\overline{\mathcal{S}}^3}\right]\frac{N_{tel}^4}{\rho_0^3\overline{K}^3}
+
\left[\frac{3N\sigma^4}{(1-\tau)^4V_{\star}^2(f_{ij})V_{\star}^2(f_{kl})\mathcal{S}^4}\right]\frac{N_{tel}^4}{\rho_0^4\overline{K}^4}$\\
&coefficients&$\ds {\mathrm{if}~~j=k} +
\frac{1}{\tau\overline{\mathcal{S}}}\frac{N_{tel}}{\rho_0\overline{K}}
+ \left[\frac{2\mathrm{Re}[V_{\star}(f_{ij})V_{\star}(f_{kl})V^{\ast}_{\star}(f_{il})]}{(1-\tau)V_{\star}^2(f_{ij})V_{\star}^2(f_{kl})\overline{\mathcal{S}}}\right]\frac{N_{tel}}{\rho_0\overline{K}}
 +
 \left[\frac{V_{\star}^2(f_{il})}{(1-\tau)^2V_{\star}^2(f_{ij})V_{\star}^2(f_{kl})\mathcal{S}^2}\right]\frac{N_{tel}^2}{\rho_0^2\overline{K}^2} 
$ \\ &&\\ \hline
\end{tabular}
}
\end{sidewaystable}

\clearpage

\begin{sidewaystable}
\centering
\caption{\label{table_covclotphot}Description of the elements
    of the closure phase covariance matrix, in the ``photon noise"
    regime.}
{\scriptsize
\begin{tabular}{|c|c|l|}
\hline
&  \multicolumn{2}{|c|}{}\\
Observables &  \multicolumn{2}{|c|}{Covariance coefficients}  \\
&  \multicolumn{2}{|c|}{}\\
\hline 
&&\\
 &  & $\ds \left[\frac{[N_{tel}^3-
   2V_{\star}^{\ast}(f_{ij})V_{\star}^{\ast}(f_{jk})V_{\star}(f_{ik})]
   \overline{\mathcal{S}}^3 +
 3N_{tel}^2\sigma^2_{\mathcal{S}}\overline{\mathcal{S}}}{2|V_{\star}(f_{ij})|^2|V_{\star}(f_{jk})|^2|V_{\star}(f_{ik})|^2\overline{\mathcal{S}}^6}\right]\frac{N_{tel}^3}{\rho_0^3\overline{K}^3} 
+
\left[\frac{[|V_{\star}(f_{ij})|^2 +
 |V_{\star}(f_{jk})|^2 +
 |V_{\star}(f_{ik})|^2][N_{tel}^2\overline{\mathcal{S}}^4 +
 N_{tel}\sigma^2_{\mathcal{S}}\overline{\mathcal{S}}^2 +
 2\sigma^4_{\mathcal{S}}]}{2|V_{\star}(f_{ij})|^2|V_{\star}(f_{jk})|^2|V_{\star}(f_{ik})|^2\overline{\mathcal{S}}^6}\right]\frac{N_{tel}^2}{\rho_0^2\overline{K}^2}$
  \\
 &  Diagonal & $\ds -\left[\frac{[|V_{\star}(f_{ij})|^4 +
 |V_{\star}(f_{jk})|^4 +
 |V_{\star}(f_{ik})|^4][\overline{\mathcal{S}}^4 +
 2\sigma^2_{\mathcal{S}}\overline{\mathcal{S}}^2
 +\sigma^4_{\mathcal{S}}] + 2[|V_{\star}(f_{ij})|^2
 |V_{\star}(f_{jk})|^2 + |V_{\star}(f_{ik})|^2 (|V_{\star}(f_{jk})|^2+
 |V_{\star}(f_{ij})|^2)][\overline{\mathcal{S}}^4 + \sigma^2_{\mathcal{S}}\overline{\mathcal{S}}^2]}{2|V_{\star}(f_{ij})|^2|V_{\star}(f_{jk})|^2|V_{\star}(f_{ik})|^2\overline{\mathcal{S}}^6}\right]\frac{N_{tel}^2}{\rho_0^2\overline{K}^2}$
\\
&coefficients& $\ds + \left[\frac{[|V_{\star}(f_{ij})|^2
 |V_{\star}(f_{jk})|^2 + |V_{\star}(f_{ik})|^2 (|V_{\star}(f_{jk})|^2+
 |V_{\star}(f_{ij})|^2)][N_{tel}\overline{\mathcal{S}}^5 +
 (N_{tel}+4)\sigma^2_{\mathcal{S}}\overline{\mathcal{S}}^3 +
 2\sigma^4_{\mathcal{S}}\overline{\mathcal{S}}]}{2|V_{\star}(f_{ij})|^2|V_{\star}(f_{jk})|^2|V_{\star}(f_{ik})|^2\overline{\mathcal{S}}^6}\right]\frac{N_{tel}}{\rho_0\overline{K}} $\\
Closure phase &&$\ds -\left[\frac{2[V_{\star}(f_{ij})V_{\star}(f_{jk})V_{\star}^{\ast}(f_{ik})]
 [|V_{\star}(f_{ij})|^2 +
 |V_{\star}(f_{jk})|^2 +
 |V_{\star}(f_{ik})|^2][\overline{\mathcal{S}}^5 + 2\sigma^2_{\mathcal{S}}\overline{\mathcal{S}}^3 +
 \sigma^4_{\mathcal{S}}\mathcal{S}]}{2|V_{\star}(f_{ij})|^2|V_{\star}(f_{jk})|^2|V_{\star}(f_{ik})|^2\overline{\mathcal{S}}^6}\right]\frac{N_{tel}}{\rho_0\overline{K}}$ \\
&&\\
\cline{2-3}
&&\\
&& $0~~~$ if $f_{ij} \neq f_{lm}$ \\
$\ds \mathrm{Cov}\left\{\Phi_{ijk},\Phi_{lmn}\right\}$ &&\\
& &   
$\ds 
\left[\frac{|V_{\star}(f_{kn})|^2[N_{tel}\overline{\mathcal{S}}^3 +
  2\sigma^2_{\mathcal{S}}\overline{\mathcal{S}}]}{2V_{\star}(f_{ij})V_{\star}(f_{jk})|V_{\star}^{\ast}(f_{ik})V^{\ast}_{\star}(f_{lm})V^{\ast}_{\star}(f_{mn})|V_{\star}(f_{ln})\overline{\mathcal{S}}^6}\right] \frac{N_{tel}^3}{\rho_0^3\overline{K}^3}$ \\
& Non diagonal & $\ds + \left[\frac{[|V_{\star}(f_{ij})|^2|V_{\star}(f_{kn})|^2]\overline{\mathcal{S}}^4  +
[V_{\star}(f_{kn})V_{\star}^{\ast}(f_{mn})V_{\star}^{\ast}(f_{mk})
+
V_{\star}(f_{kn})V_{\star}^{\ast}(f_{lk})V_{\star}^{\ast}(f_{ln})][N_{tel}\overline{\mathcal{S}}^4 +
  3\sigma^2_{\mathcal{S}}\overline{\mathcal{S}}^2]}{2V_{\star}(f_{ij})V_{\star}(f_{jk})|V_{\star}^{\ast}(f_{ik})V^{\ast}_{\star}(f_{lm})V^{\ast}_{\star}(f_{mn})|V_{\star}(f_{ln})\overline{\mathcal{S}}^6}\right]\frac{N_{tel}^2}{\rho_0^2\overline{K}^2}$\\
& coefficients &$\ds -\left[\frac{[|V_{\star}(f_{ik})|^2|V_{\star}(f_{in})|^2
  +
  |V_{\star}(f_{jk})|^2|V_{\star}(f_{jn})|^2][\overline{\mathcal{S}}^4+\sigma^2_{\mathcal{S}}\overline{\mathcal{S}}^2] + [|V_{\star}(f_{ik})|^2|V_{\star}(f_{jn})|^2 +
  |V_{\star}(f_{in})|^2|V_{\star}(f_{jk})|^2]}{2V_{\star}(f_{ij})V_{\star}(f_{jk})|V_{\star}^{\ast}(f_{ik})V^{\ast}_{\star}(f_{lm})V^{\ast}_{\star}(f_{mn})|V_{\star}(f_{ln})\overline{\mathcal{S}}^6}\right]\overline{\mathcal{S}}^4\frac{N_{tel}^2}{\rho_0^2\overline{K}^2}$ \\ 
&& $\ds \left[\frac{[V_{\star}(f_{jk})V_{\star}^{\ast}(f_{ik})V_{\star}^{\ast}(f_{jn})V_{\star}(f_{in})][N_{tel}\overline{\mathcal{S}}^5 +
  4\sigma^2_{\mathcal{S}}\overline{\mathcal{S}}^3]+[V_{\star}(f_{kn})V_{\star}^{\ast}(f_{in})V_{\star}^{\ast}(f_{ik})
+ V_{\star}(f_{kn})V_{\star}(f_{jk})V_{\star}^{\ast}(f_{jn})]|V_{\star}(f_{ij})|^2[\overline{\mathcal{S}}^5 +
  \sigma^2_{\mathcal{S}}\overline{\mathcal{S}}^3]}{2V_{\star}(f_{ij})V_{\star}(f_{jk})|V_{\star}^{\ast}(f_{ik})V^{\ast}_{\star}(f_{lm})V^{\ast}_{\star}(f_{mn})|V_{\star}(f_{ln})\overline{\mathcal{S}}^6}\right]\frac{N_{tel}}{\rho_0\overline{K}}$\\
&& $\ds -\left[\frac{[V_{\star}(f_{ij})V_{\star}(f_{jk})V_{\star}^{\ast}(f_{ik})(|V_{\star}(f_{in})|^2+|V_{\star}(f_{jn})|^2)    +
  V_{\star}(f_{ij})V_{\star}(f_{jn})V_{\star}^{\ast}(f_{in})(|V_{\star}(f_{jk})|^2+|V_{\star}(f_{ik})|^2)][\overline{\mathcal{S}}^5+\sigma^2_{\mathcal{S}}\overline{\mathcal{S}}^3]}{2V_{\star}(f_{ij})V_{\star}(f_{jk})|V_{\star}^{\ast}(f_{ik})V^{\ast}_{\star}(f_{lm})V^{\ast}_{\star}(f_{mn})|V_{\star}(f_{ln})\overline{\mathcal{S}}^6}\right]\frac{N_{tel}}{\rho_0\overline{K}}~~~$ if $f_{ij} = f_{lm}$ \\
&&\\ \hline
\end{tabular}
}
\end{sidewaystable}

\begin{sidewaystable}
\centering
\caption{\label{table_covclotdet}Description of the elements
    of the closure phase covariance matrix, in the ``detector noise"
    regime.}
{\scriptsize
\begin{tabular}{|c|c|l|}
\hline
&  \multicolumn{2}{|c|}{}\\
Observables &  \multicolumn{2}{|c|}{Covariance coefficients}  \\
&  \multicolumn{2}{|c|}{}\\
\hline
&&\\
 & Diagonal  & $\ds \left[\frac{[|V_{\star}(f_{ij})|^2+|V_{\star}(f_{jk})|^2+|V_{\star}(f_{ik})|^2][3N\sigma^4+N^2\sigma^4]\overline{\mathcal{S}}^2}{2|V_{\star}(f_{ij})|^2|V_{\star}(f_{jk})|^2|V_{\star}(f_{ik})|^2\overline{\mathcal{S}}^6}\right]\frac{N_{tel}^4}{\rho_0^4\overline{K}^4}
$
  \\
Closure phase & coefficients& $\ds +\left[\frac{[|V_{\star}(f_{ij})|^2|V_{\star}(f_{ij})|^2+|V_{\star}(f_{jk})|^2|V_{\star}(f_{jk})|^2+|V_{\star}(f_{ik})|^2|V_{\star}(f_{ij})|^2]N\sigma^2[\overline{\mathcal{S}}^4
    +
 \sigma^2_{\mathcal{S}}\overline{\mathcal{S}}^2]}{2|V_{\star}(f_{ij})|^2|V_{\star}(f_{jk})|^2|V_{\star}(f_{ik})|^2\overline{\mathcal{S}}^6}\right]\frac{N_{tel}^2}{\rho_0^2\overline{K}^2} + N^3\sigma^6+3N^2\sigma^6$
\\
&&\\
\cline{2-3}
&&\\
&& $0~~~$ if $f_{ij} \neq f_{lm}$\\
$\ds \mathrm{Cov}\left\{\Phi_{ijk},\Phi_{lmn}\right\}$ &Non diagonal&\\
& coefficients  &   
$\ds
\left[\frac{[V_{\star}(f_{jk})V_{\star}^{\ast}(f_{ik})V_{\star}^{\ast}(f_{jn})V_{\star}(f_{in})]N\sigma^2\overline{\mathcal{S}}^4}{2V_{\star}(f_{ij})V_{\star}(f_{jk})|V_{\star}^{\ast}(f_{ik})V^{\ast}_{\star}(f_{lm})V^{\ast}_{\star}(f_{mn})|V_{\star}(f_{ln})\overline{\mathcal{S}}^6}\right]\frac{N_{tel}^2}{\rho_0^2\overline{K}^2}
+ 3N^2\sigma^6~~~$ if $f_{ij} = f_{lm}$ \\
&& $\ds $\\
&&\\ \hline
\end{tabular}
}
\end{sidewaystable}


\clearpage

\listoffigures

\clearpage
 
\clearpage

\clearpage

\clearpage

\clearpage

\clearpage

\clearpage

\end{document}